\documentclass[12pt]{article}
\usepackage{axodraw}
\usepackage{epsf}

\def\theequation{\arabic{section}.\arabic{equation}}

\begin{document}
~\vspace{-1.cm}
\begin{flushright}
ANL-HEP-PR-01-112\\[-0.15cm]
CERN-TH/2001-307\\[-0.15cm]
EFI-2001-052\\[-0.15cm]
FERMILAB-Pub-01/353-T\\[-0.15cm]
MC-TH-2001-011\\[-0.15cm]
hep-ph/0111245\\[-0.15cm]
November 2001
\end{flushright}   

\begin{center}
{\LARGE {\bf Higgs-Boson Pole Masses in the MSSM with}}\\[0.3cm]
{\LARGE {\bf  Explicit CP Violation}}\\[1.4cm]
{\large M. Carena$^{\,a}$, J. Ellis$^{\, b}$, 
A. Pilaftsis$^{\, c}$ and C.E.M. Wagner$^{\,d,e}$}\\[0.4cm]
$^a${\em Fermilab, P.O. Box 500, Batavia IL 60510, U.S.A.}\\[0.2cm]
$^b${\em Theory Division, CERN, CH-1211 Geneva 23, Switzerland}\\[0.2cm]
$^c${\em Department of Physics and Astronomy, University of Manchester,\\
Manchester M13 9PL, United Kingdom}\\[0.2cm]
$^d${\em High Energy Physics Division, Argonne National Lab., Argonne
  IL 60439, U.S.A.}\\[0.2cm]
$^e${\em Enrico Fermi Institute, University of Chicago, 5640 Ellis Ave.,
 Chicago IL 60637, U.S.A.}
\end{center}
\vskip1.cm   \centerline{\bf  ABSTRACT}  
Extending  previous   results  obtained  in   the  effective-potential
approach, we  derive analytic expressions  for the pole masses  of the
charged  and  neutral  Higgs  bosons  in  the  minimal  supersymmetric
standard model (MSSM) with explicit CP violation.  In such a
minimal supersymmetric model, the CP invariance of the Higgs potential
is explicitly broken by  quantum effects dominated by third-generation
squarks   at  the  one-loop   level  and   by  one-loop   gluino-  and
higgsino-mediated threshold  corrections to the  top- and bottom-quark
Yukawa  couplings  at  the  two-loop  level.   Field-theoretic  issues
arising in the description  of a two-level Higgs-boson system strongly
mixed due  to CP violation  are addressed.  Numerical results  for the
pole  masses  of the  CP-violating  Higgs  bosons  are presented,  and
comparisons   are  made   with  previous   results  obtained   in  the
effective-potential approach.

\newpage

\setcounter{equation}{0}
\section{Introduction}

The minimal supersymmetric extension of the Standard Model (MSSM)
represents the most economical, and perhaps the most appealing,
low-energy realization of supersymmetry (SUSY) softly broken at
0.1--1~TeV energies, within which a natural formulation of the
gauge-hierarchy problem is possible~\cite{HPN}.  Unlike the Standard
Model (SM), the MSSM exhibits gauge-coupling unification at very high
energy scales of order $10^{16}$~GeV~\cite{DG}, offers a theoretical
framework able to realize cosmological baryogenesis via a strongly
first-order electroweak phase transition~\cite{KRS,EWBAU}, and
provides an interesting candidate for the cold dark matter favoured by
astrophysics~\cite{DM}. Moreover, an important prediction of the MSSM
is the existence of a light neutral Higgs boson which, in general, has
SM-like properties and whose mass cannot exceed a calculable upper
bound of about 130~GeV~\cite{Mh}.  Since supersymmetric particle
contributions to the precision electroweak observables rapidly
decouple for values of the sparticle masses above the weak scale,
these values of the lightest neutral Higgs-boson mass are consistent
with the analysis of precision electroweak data, which, within the
Standard Model, lead to a preference for a relatively Higgs-boson
mass, not larger than $\sim 240$~GeV~\cite{LEPEWWG}.

It has  been emphasized in many papers  that loop effects dominated by
third-generation squarks may   violate   sizeably the tree-level   CP
invariance                   of          the       MSSM          Higgs
potential~\cite{APLB,PW,Demir,CDL,CEPW,KW,CHL,INrecent,CPX,Ibr,HeinCP},    
giving
rise  to important modifications of  the Higgs-boson  couplings to the
$W^\pm$ and $Z$ bosons and to fermionic matter. Radiative CP violation
in the MSSM Higgs sector  thereby affects significantly the production
rates   of the Higgs  bosons.   In  particular, modifications of   the
couplings to  the $Z$ boson of the  lightest and second-lightest Higgs
bosons might  allow a relatively  light  Higgs  boson with a  mass  of
60--70~GeV  to  have  escaped  detection   at LEP~2~\cite{CEPW}.   The
upgraded Tevatron collider,  the LHC and  a linear  $e^+ e^-$ collider
will  be able to probe in detail   
this minimal CP-violating supersymmetric model.

An accurate determination of the MSSM Higgs-boson mass spectrum in the
presence of CP violation plays an important r\^ole in the studies of
electroweak baryogenesis~\cite{EWBAU}.  It may also be important for
understanding the domain of MSSM parameter space compatible with
supersymmetric dark matter~\cite{DM}, since CP violation is known to
extend the allowed ranges of parameters, and even in the CP-conserving
case rapid annihilation of relic particles via direct-channel Higgs
poles can yield `funnels' of parameter space extending to
exceptionally large sparticle masses.

In
earlier work, we computed the  charged and neutral Higgs-boson  masses
from  the  one-loop    effective  MSSM  Higgs     potential, utilizing
renormalization-group  (RG)  methods.   Here,  we  extend  our earlier
computation  of RG-improved Higgs-boson  masses by taking into account
finite contributions  that arise from  the shifts in the  positions of
the poles in  the corresponding Higgs-boson propagators.  Even  though
the   computation  of  the Higgs-boson pole    masses that  we make is
diagrammatic at the one-loop level, we are still able to implement the
RG techniques known  from previous effective-potential analyses in the
MSSM~\cite{KYS,CEQR,CEQW,CQW}.  In  fact,  one  can establish a  close
relationship   between    the RG-improved effective-potential approach
followed in~\cite{CEPW}  and  the RG-improved diagrammatic calculation
presented here. Specifically, if one considers the zero-momentum limit
of the Higgs-boson propagators  within the diagrammatic framework, the
two approaches give identical results, which constitutes a non-trivial
check on the correctness of our numerical evaluations.

Following~\cite{CEPW},   we include  in  our  RG-improved diagrammatic
approach  the leading two-loop logarithms due   to QCD corrections, as
well as  the  leading  two-loop  logarithms induced  by  the top-  and
bottom-quark Yukawa couplings~\cite{CQW}.   In addition, we take  into
consideration the   leading  one-loop logarithms   due to  gaugino and
higgsino quantum effects~\cite{HH}.   Finally, we add to the effective
potential  the  potentially large two-loop non-logarithmic corrections
originating  from one-loop    threshold   effects on   the    top- and
bottom-quark Yukawa couplings, associated  with the decoupling of  the
third-generation squarks~\cite{CHHHWW}.

As we show below, sparticle loops may have a relevant impact
on the definition of the pole masses. Although the shift of the
lightest neutral Higgs-boson mass is typically of the order of a few GeV,
and therefore of the order of the uncertainties implicit in the effective
potential computation of the neutral Higgs masses, the shift of
the heavier neutral and charged Higgs-boson masses may be much larger,
of the order of a few tens of GeV. These relatively large differences
between the pole and running masses are usually associated with 
Higgs-boson mass values which are close to the thresholds for 
on-shell production of the third-generation squarks, and may have
important consequences for Higgs physics at future colliders.

The organization of the paper is as follows. Section~2 describes the
approach we use to evaluate the Higgs-boson pole masses.  Based on this
approach, we compute in Section~3 the pole masses of the charged and
neutral Higgs bosons. Analytic expressions useful in the computation are
given in Appendices A and B. We also address in Section~3 some crucial
field-theoretic issues related to the proper description of a two-level
Higgs-boson system strongly mixed via CP violation. Section~4 presents
numerical estimates of Higgs-boson pole masses, making comparisons with
previous results obtained in the RG-improved effective-potential approach.
Our conclusions are summarized in Section~5~\cite{subroutine}.

\setcounter{equation}{0}
\section{RG-Improved Higgs-Boson Self-Energies}

In this Section, we describe our approach to calculating the pole masses
of the charged and neutral Higgs bosons in the MSSM with explicit
radiative CP violation. Our approach utilizes RG methods in the
$\overline{\rm MS}$ scheme developed earlier in~\cite{KYS,CEQR,CEQW,CQW}.
In particular, we demonstrate that, in the limit in which the squared
off-shell momentum $s=p^2$ of the Higgs-boson propagators goes to zero,
our derived analytic results of the Higgs-boson masses are identical to
those previously obtained in~\cite{CEPW} in the effective Higgs-potential
approach.

First we briefly review the basic low-energy structure of the MSSM
with explicit CP violation. It is possible for CP violation to appear
via the Higgs superpotential and the soft SUSY-breaking Lagrangian:
\begin{eqnarray}
  \label{Wpot}
W & = & h_l\, \widehat{H}^T_1 i\tau_2 \widehat{L} \widehat{E}\: +\: 
h_d\, \widehat{H}^T_1 i\tau_2 \widehat{Q} \widehat{D}\: +\: 
h_u\, \widehat{Q}^T i\tau_2 \widehat{H}_2 \widehat{U}\: -\:
\mu\,\widehat{H}^T_1 i\tau_2 \widehat{H}_2\, ,\\
  \label{Lsoft}
-{\cal L}_{\rm soft} & = & 
-\, \frac{1}{2}\, \Big( m_{\tilde{g}}\,
\lambda^a_{\tilde{g}}\lambda^a_{\tilde{g}}\:+\: 
m_{\widetilde{W}}\,\lambda^i_{\widetilde{W}}\lambda^i_{\widetilde{W}}\: 
+\: m_{\widetilde{B}}\,\lambda_{\widetilde{B}}\lambda_{\widetilde{B}}
\: +\: {\rm h.c.}\Big)\:
+\: \widetilde{M}^2_L\, \widetilde{L}^\dagger 
\widetilde{L}\: +\: \widetilde{M}^2_Q\, \widetilde{Q}^\dagger 
\widetilde{Q}\nonumber\\
&& +\, \widetilde{M}^2_U\,\widetilde{U}^*\widetilde{U}\: 
+\: \widetilde{M}^2_D\,\widetilde{D}^* \widetilde{D}\: 
+\: \widetilde{M}^2_E\, \widetilde{E}^* \widetilde{E}\: +\: 
m^2_1\, \widetilde{\Phi}^\dagger_1\widetilde{\Phi}_1\: +\:
m^2_2\, \Phi^\dagger_2\Phi_2\: -\: \Big(m^2_{12}\,\widetilde{\Phi}^T_1 i\tau_2
\Phi_2\nonumber\\
&& +\: {\rm h.c.}\Big)\: 
+\: \Big(\, h_l A_l\,\Phi_1^\dagger \widetilde{L} \widetilde{E}\: +\:
h_dA_d\,\Phi_1^\dagger \widetilde{Q} \widetilde{D}\: -\:  
h_uA_u\,\Phi_2^T i\tau_2 \widetilde{Q} \widetilde{U}\ +\ {\rm h.c.}\, \Big)\,,
\end{eqnarray}
where $\widetilde{\Phi}_1 = i\tau_2  \Phi^*_1$ is the physical bosonic
degree of freedom of the  Higgs chiral superfield $\widehat{H}_1$, and
$\tau_2$  is the usual Pauli matrix.   Throughout  the paper, we
follow  the  conventions  of~\cite{PW,CEPW}.  Details of the notation,
including  the quantum-number assignments of the fields under the SM
gauge group, are given in Table~1 of~\cite{CEPW}.

The expressions (\ref{Wpot}) and (\ref{Lsoft}) contain a large number
of new CP-violating phases in the MSSM that are not present in the
SM. The large number of these independent CP-odd phases can be
substantially reduced if universality conditions are imposed on the
complex soft SUSY-breaking parameters of the theory at some
high-energy scale $M_X$.  Under such universality conditions, all soft
SUSY-breaking squark masses are set equal to a universal value at the
scale $M_X$. Similarly, all the soft SUSY-breaking Yukawa coupling
coefficients $A_f$ are set to a universal value at $M_X$, and likewise
the gaugino masses $m_{\tilde{g}}$, $m_{\widetilde{W}}$ and
$m_{\widetilde{B}}$ of the gauge groups SU(3)$_c$, SU(2)$_L$ and
U(1)$_Y$ are assumed to share a common value $m_{1/2}$ at $M_X$.  In
this unified framework~\footnote[1]{One should bear in mind that the
different RG runnings of the phases of the soft SUSY-breaking Yukawa
couplings and of the gaugino masses lead to non-universal values at
lower energies~\cite{Wells}.}, the new CP-odd phases are contained in
four complex parameters: the $\mu$ parameter that mixes the Higgs
supermultiplets, the soft bilinear Higgs-mixing mass-squared parameter
$m^2_{12}$, the common gaugino mass $m_{1/2}$ and the common soft
Yukawa coupling coefficient $A_f$.  However, two of these four phases
may be eliminated using the two global symmetries that govern the
dimension-four operators in the MSSM Lagrangian.  In this way, the
parameters $\mu$ and $m^2_{12}$ may be rephased so as to become real
numbers. Consequently, the physical CP-violating phases of interest
are ${\rm arg}\, (A_f)$ and ${\rm arg}\, (m_{1/2})$. In this work we
shall give general expressions, which may be applicable, but are not
restricted, to the case of universal boundary conditions.

The ground state of the radiatively-corrected MSSM Higgs potential
$-{\cal L}_V$ may be determined by first  linearly expanding the Higgs
doublets $\Phi_1$ and $\Phi_2$ as
\begin{equation}
  \label{Phi12}
\Phi_1\ =\ \left( \begin{array}{c}
\phi^+_1 \\ \frac{1}{\sqrt{2}}\, ( v_1\, +\, \phi_1\, +\, ia_1)
\end{array} \right)\, ,\qquad
\Phi_2\ =\ e^{i\xi}\, \left( \begin{array}{c}
\phi^+_2 \\  \frac{1}{\sqrt{2}}\, ( v_2 \, +\, \phi_2\, +\, ia_2 )
 \end{array} \right)\, ,
\end{equation}
where $v_1$ and $v_2 e^{i\xi}$ are the VEVs of the Higgs doublets, and
then demanding that  the  total tadpole  contributions to the  neutral
CP-even    and     CP-odd    fields   $\phi_{1,2}$     and   $a_{1,2}$
vanish~\cite{APLB,PW,CEPW}.  If $m^2_{12}$ is real  at the tree level,
then a relative phase $\xi$ of the  two VEVs is formally induced at
the  one-loop level.  However,  the phase  $\xi$ can always  be set to
zero  by a   judicious  choice of   the counter-term  (CT)  ${\rm  
Im}~m^2_{12}$.  Specifically, the CT ${\rm Im}~m^2_{12}$  can be chosen in
a way such that it cancels completely the CP-odd tadpole graphs, leading
to  the consistent solution $\xi  = 0$ for the  vanishing of the total
CP-odd  tadpole  contributions. In this context, we
stress again~\cite{APLB} that such a $\xi = 0$ scheme of
renormalization preserves the   phase  convention for $m^2_{12}$, and
consequently for $\mu$, beyond  the Born approximation. It may be extended
straightforwardly to all orders in perturbation theory.

It is often more convenient to switch to a basis in which the would-be
Goldstone bosons $G^\pm$ and $G^0$ of the $W^\pm$ and $Z$ gauge bosons
decouple at  $s = p^2 = 0$ from the  charged Higgs-boson and pseudoscalar 
mass
matrices, respectively.  In such a basis, the  new Higgs fields may be
expressed in terms of the old ones through the orthogonal rotations:
\begin{equation}
  \label{rot}
\left(\! \begin{array}{c} G^+ \\ H^+ \end{array}\!\right)\ = \ 
\left(\! \begin{array}{cc} \cos\beta & \sin\beta \\ 
-\,\sin\beta & \cos\beta \end{array}\!\right)\, 
\left(\! \begin{array}{c} \phi^+_1 \\ \phi^+_2 \end{array}\!\right)\,,\qquad
\left(\! \begin{array}{c} G^0 \\ a \end{array}\!\right)\ = \ 
\left(\! \begin{array}{cc} \cos\beta & \sin\beta \\ 
-\,\sin\beta & \cos\beta \end{array}\!\right)\, \left(\! \begin{array}{c}
a_1 \\ a_2 \end{array}\!\right) \,,
\end{equation}
where $\tan\beta \equiv v_2/v_1$. In~(\ref{rot}), $H^+$ and $a$ are
the charged and `CP-odd' neutral Higgs scalars in the effective
potential limit: $s\to 0$.  As we see explicitly in the next Section,
the effective masses of the physical Higgs-boson fields found in the
limit $s\to 0$ differ from the physical ones determined by the poles
of the general propagator matrices.


\begin{figure}[t]
\begin{center}
\begin{picture}(400,200)(0,0)
\SetWidth{0.8}


\DashArrowLine(10,150)(50,150){4}\Text(40,160)[r]{$\phi_j,a_j,\phi^-_j$}
\DashArrowLine(100,150)(140,150){4}\Text(110,160)[l]{$\phi_i,a_i,\phi^-_i$}
\DashArrowArc(75,150)(25,0,180){5}\Text(75,185)[]{$\tilde{t}_{1,2},
\tilde{b}_{1,2}$}\DashArrowArc(75,150)(25,180,360){5}

\Text(75,110)[]{\bf (a)}


\DashArrowLine(210,125)(260,125){4}\Text(235,135)[r]{$\phi_j,a_j,\phi^-_j$}
\DashArrowLine(260,125)(310,125){4}\Text(285,135)[l]{$\phi_i,a_i,\phi^-_i$}
\DashArrowArc(260,150)(25,0,360){5}\Text(260,185)[]{$\tilde{t}_{1,2},
\tilde{b}_{1,2}$}

\Text(260,110)[]{\bf (b)}


\DashArrowLine(10,50)(50,50){4}\Text(40,60)[r]{$\phi_j,a_j,\phi^-_j$}
\DashArrowLine(100,50)(140,50){4}\Text(110,60)[l]{$\phi_i,a_i,\phi^-_i$}
\ArrowArc(75,50)(25,0,180)\Text(75,85)[]{$t,b$}
\ArrowArc(75,50)(25,180,360)

\Text(75,10)[]{\bf (c)}

\DashArrowLine(200,20)(200,50){4}\Text(195,25)[r]{$\phi_i,a_i$}
\DashArrowArc(200,70)(20,0,360){5}\Text(200,100)[]{$\tilde{t}_{1,2},
\tilde{b}_{1,2}$}

\Text(200,10)[]{\bf (d)}

\DashArrowLine(320,20)(320,50){4}\Text(315,25)[r]{$\phi_i$}
\ArrowArc(320,70)(20,0,360)\Text(320,100)[]{$t,b$}

\Text(320,10)[]{\bf (e)}

\end{picture}
\end{center}
\caption{Third-generation quark and squark contributions to the Higgs-boson 
self-energies.}\label{f1}
\end{figure}
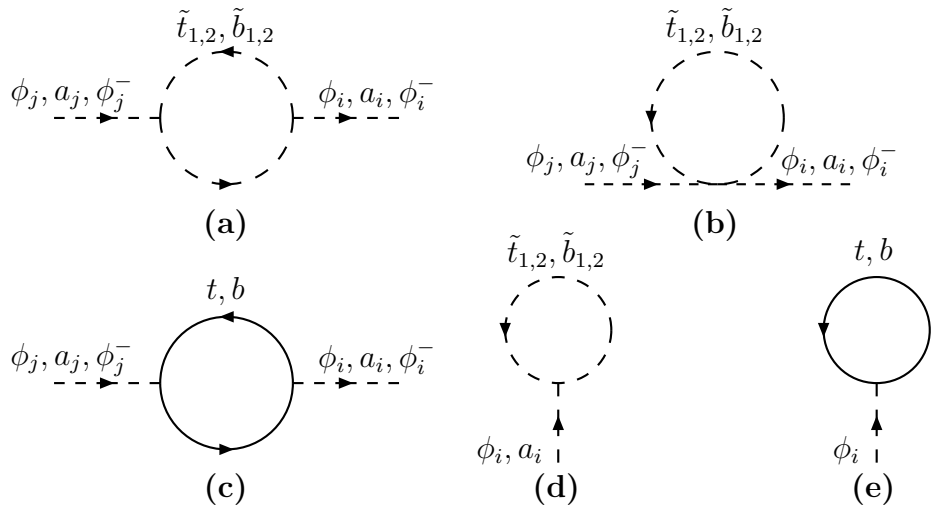

We first describe our programme for renormalizing the charged
Higgs-boson self-energies; extending this programme to the
neutral Higgs-boson  self-energies is then straightforward.  We
adopt  the $\overline{\rm MS}$ scheme  of  renormalization.  The
charged Higgs-boson self-energy  transitions $\phi^-_j  \to \phi^-_i$,
with  $i,j  =1,2$,  span a   $2\times  2$-dimensional matrix, which we
denote by $\Pi^{\pm}(s)$.   As  shown in Fig.~\ref{f1},  the  dominant
contributions to  $\Pi^{\pm}(s)$ come from quarks  and  squarks of the
third   generation.  Employing the  couplings  of  the Higgs bosons to
squarks listed   in  Appendix A,  we  derive  in  Appendix  B analytic
expressions for the  unrenormalized  charged Higgs-boson self-energies
$\Pi^{\pm}(s)$.   Then,  in the weak  basis $\{\phi^+_1,\ \phi^+_2\}$,
the corresponding renormalized self-energies, denoted by
$\widetilde{\Pi}^{\pm}(s)$, are given by
\begin{equation}
  \label{Pch}
\widetilde{\Pi}^{\pm}(s)\ =\ 
\Pi^{\pm,\overline{\rm MS}}(s)\ +\ \left(\! \begin{array}{cc}
\frac{\displaystyle T^{\overline{\rm MS}}_{\phi_1}}{\displaystyle v_1}  &  
i\, \frac{\displaystyle T^{\overline{\rm MS}}_a}{\displaystyle v}\\[0.3cm]
-\,i\,\frac{\displaystyle T^{\overline{\rm MS}}_a}{\displaystyle v}  &  
\frac{\displaystyle T^{\overline{\rm MS}}_{\phi_2}}{\displaystyle v_2}
\end{array}\!\right)\,,
\end{equation}
with  $T_a =  T_{a_2}/\cos\beta    = -   T_{a_1}/\sin\beta$.  In   the
$\overline{\rm   MS}$ scheme, the   renormalized  charged  Higgs-boson
self-energy   $\Pi^{\pm,\overline{\rm   MS}}(s)$  is    obtained  from
$\Pi^{\pm}(s)$ by  simply dropping  all  terms proportional to  the 
ultra-violet (UV)
divergence  $C_{\rm   UV}$.  Such  a    scheme  of renormalization  is
valid for renormalizable theories, such as the one 
under study. Similarly, in~(\ref{Pch}), $T^{\overline{\rm
MS}}_{\phi_{1,2}}$  and  $T^{\overline{\rm   MS}}_{a_{1,2}}$  are  the
$\overline{\rm   MS}$-renormalized   tadpole  CTs    as    derived from
$T_{\phi_{1,2}}$  and $T_{a_{1,2}}$.  The    latter are  computed   in
Appendix B.

We can now extend the procedure for the RG improvement of effective
Higgs-boson mass matrices we presented in~\cite{CEPW} to Higgs-boson
self-energies.  In particular, we wish to take properly into account
the decoupling properties and threshold effects of the heavy squarks
in the loops. Following a line of argument closely related to that
in~\cite{CEPW}, we arrive at the RG-improved charged Higgs-boson
self-energy matrix
\begin{equation}
  \label{RGplus}
\widehat{\Pi}^\pm_{ij}(s)\ =\ -\,(\overline{\cal M}^2_{\pm})_{ij}\: +\:
(\xi^+_i \xi^-_j)^{-1}\, (\Delta \Pi^\pm)^{\tilde{t}\tilde{b}}_{ij}(s)\: 
+\: (\widetilde{\Pi}^\pm)^{tb}_{ij} (s)\,.
\end{equation}
Here,  $\overline{\cal M}^2_{\pm}$ is  the two-loop Born-improved mass
matrix given by
\begin{equation}
  \label{Mbarplus}
\overline{\cal M}^2_{\pm}\ =\ {\cal M}^{2\,(0)}_{\pm} 
\Big(\,{\rm Re}\, \bar{m}^2_{12}\,,\ \bar{\lambda}_4\, \Big)\,, 
\end{equation}
where 
\begin{equation}
  \label{M0plus}
{\cal M}^{2\,(0)}_{\pm}\Big(\,{\rm Re}\, m^2_{12}\,,\ \lambda_4\, \Big) \ =\ 
\bigg(\, \frac{1}2\, \lambda_4\, v_1 v_2\: +\: {\rm Re}\, m^2_{12}\, \bigg)\,
\left(\! \begin{array}{cc}
\tan\beta & -1 \\ -1 & \cot\beta \end{array}\!\right)
\end{equation}
is the tree-level charged  Higgs-boson mass matrix expressed  in terms
of relevant parameters such  as  the real part of  the
soft bilinear   Higgs mixing, ${\rm  Re}\,  m^2_{12}$, and the quartic
coupling~$\lambda_4$. The bar on these parameters 
in~(\ref{Mbarplus}), i.e.\ ${\rm Re}\, \bar{m}^2_{12}$  and
$\bar{\lambda}_4$, indicates the sum of the tree-level and of the one-
and two-loop leading logarithmic contributions.  The analytic forms of
${\rm Re}\, \bar{m}^2_{12}$ and $\bar{\lambda}_1\dots \bar{\lambda}_4$
are given in~\cite{CEPW}.

The second term in~(\ref{RGplus})  describes the threshold effect
of  the top and  bottom squarks and is the  product of two quantities:
(i)~the anomalous-dimension factors of the external Higgs fields
\begin{equation}
  \label{xiplus}
\xi^\pm_1\ =\ \xi^\pm_1 (m_t)\ =\ 1\: +\: \frac{3|h_b|^2}{32\pi^2}\, 
\ln\frac{Q^2_{tb}}{m^2_t}\, ,\qquad
\xi^\pm_2\ =\ \xi^\pm_2 (m_t)\ =\ 1\: +\: \frac{3|h_t|^2}{32\pi^2}\, 
\ln\frac{Q^2_{tb}}{m^2_t}\, ,
\end{equation}
where   $Q^2_{tb}   =   {\rm   max}\,   (\widetilde{M}^2_Q  +  m^2_t,\
\widetilde{M}^2_t  + m^2_t,\  \widetilde{M}^2_b  + m^2_b)$ defines the
decoupling  scale of the    heavy  squarks,  and (ii)~the    threshold
contribution
\begin{equation}
  \label{Dplus}
\Delta \Pi^\pm (s)\ =\  
{\cal M}^{2\,(0)}_{\pm} \Big(\,{\rm Re}\, m^{2\,(1)}_{12}\,,\ 
\lambda^{(1)}_4\, \Big)\: +\: \widetilde{\Pi}^\pm (s)     \,,
\end{equation}
which  is manifestly scale-independent at the  one-loop level.
The superscript `(1)' attached to the  quantities ${\rm Re}\, m^2_{12}$
and $\lambda_1\, \dots\,  \lambda_4$ indicates  one-loop  leading
logarithmic contributions. Finally, the third term in~(\ref{RGplus}) 
includes the  top- and bottom-quark effects on  the
charged Higgs-boson self-energies.

Like any quantity calculated in perturbation theory, the Higgs-boson mass
spectrum is subject to an uncertainty related to unknown higher-order
contributions, which is reflected in arbitrariness associated with the
choice of renormalization scale. In making this choice, one may use as a
guiding principle the minimization of the two-loop corrections in the
effective-potential limit $s\to 0$~\cite{CEQR,CEQW}.  Since for 
sparticle masses above the weak scale
this minimization
was shown to occur close to the top-quark mass $m_t$~\cite{CEQR,CEQW}, we
perform our computations using this renormalization 
scale.
Note that in
the limit $s\to 0$, the RG-improved charged Higgs-boson mass matrix ${\cal
M}^2_\pm$ and the corresponding self-energy matrix $\widehat{\Pi}^{\pm}
(s)$ are related by ${\cal M}^2_\pm = - \widehat{\Pi}^{\pm} (s=0)$.

The neutral Higgs-boson self-energies can be computed following a path
very analogous to the charged Higgs-boson case.  First, we compute the
$\overline{\rm    MS}$-renormalized, but not RG-improved, Higgs-boson
self-energies that   involve  the  transitions  $\phi_j \to   \phi_i$,
$\phi_j   \to a_i$ and  $a_j  \to  a_i$,  with $i,j  =  1, 2$. These
$\overline{\rm MS}$-renormalized   scalar-scalar,  scalar-pseudoscalar
and  pseudoscalar-pseudoscalar transitions are given by
the two-by-two self-energy matrices:
\begin{eqnarray}
  \label{PiSren}
\widetilde{\Pi}^S (s) &=& \Pi^{S,\overline{\rm MS}}(s)\ +\
{\rm diag}\,\bigg( \frac{T^{\overline{\rm MS}}_{\phi_1}}{v_1}\,,\
\frac{T^{\overline{\rm MS}}_{\phi_2}}{v_2}\,\bigg)\,,\nonumber\\
  \label{PiSPren}
\widetilde{\Pi}^{SP} (s) &=& \Pi^{SP,\overline{\rm MS}}(s)\ +\
i\tau_2\, \frac{T^{\overline{\rm MS}}_a}{v}\, ,\nonumber\\
  \label{PiPren}
\widetilde{\Pi}^P (s) &=& \Pi^{P,\overline{\rm MS}}(s)\ +\
{\rm diag}\,\bigg( \frac{T^{\overline{\rm MS}}_{\phi_1}}{v_1}\,,\
\frac{T^{\overline{\rm MS}}_{\phi_2}}{v_2}\,\bigg)\,,
\end{eqnarray}
where the corresponding unrenormalized  self-energies  $\Pi^S  (s)$,
$\Pi^{SP} (s)$ and $\Pi^P (s)$ may be found in Appendix B.  Then, very
analogously with the  charged Higgs-boson  case~(\ref{RGplus}),
we carry out the RG improvement of the neutral Higgs-boson
self-energies  by considering the squark threshold effects,
i.e.\ ,
\begin{equation}
  \label{RGeven}
\widehat{\Pi}^S_{ij}(s)\ =\ -\,(\overline{\cal M}^2_S)_{ij}\: +\:
\sum_{\tilde{q}=\tilde{t},{b}}\, (\xi^{\tilde{q}}_i \xi^{\tilde{q}}_j)^{-1}\, 
(\Delta \Pi^S)^{\tilde{q}}_{ij}(s)\: 
+\: (\widetilde{\Pi}^S)^{tb}_{ij} (s)\,,
\end{equation}
and likewise  for $\widehat{\Pi}^{SP}(s)$  and $\widehat{\Pi}^P (s)$.  
As in the charged  Higgs-boson   case, we denote  by $\overline{\cal
  M}^2_S$  in~(\ref{RGeven})  the CP-even  two-loop Born-improved
mass matrix and by $\xi^{\tilde{q}}_i$ the anomalous-dimension factors
\begin{eqnarray}
  \label{xitb12} \xi^{\tilde{t}}_1 (m_t) \!&=&\! 1\: +\: \frac{3
|h_b|^2}{32\pi^2}\, \ln\frac{Q^2_t}{m^2_t}\ ,\qquad \xi^{\tilde{t}}_2
(m_t) \ =\ 1\: +\: \frac{3 |h_t|^2}{32\pi^2}\, \ln\frac{Q^2_t}{m^2_t}\
,\nonumber\\ \xi^{\tilde{b}}_1 (m_t) \!&=&\! 1\: +\: \frac{3
|h_b|^2}{32\pi^2}\, \ln\frac{Q^2_b}{m^2_t}\ ,\qquad \xi^{\tilde{b}}_2
(m_t) \ =\ 1\: +\: \frac{3 |h_t|^2}{32\pi^2}\, \ln\frac{Q^2_b}{m^2_t}\ ,
\end{eqnarray} 
which appropriately define the running from the squark-decoupling scales
$Q^2_t = {\rm max}\, \Big( \widetilde{M}^2_Q + m^2_t,\ \widetilde{M}^2_t +
m^2_t \Big)$ and $Q^2_b = {\rm max}\, \Big( \widetilde{M}^2_Q + m^2_b,\
\widetilde{M}^2_b + m^2_b \Big)$ down to the top-quark mass
$m_t$~\footnote{This procedure involves the approximation of identifying
the quark Yukawa couplings with the squark ones even below the
SUSY-breaking scale.  An explicit comparison of the effective potential
computation with the diagrammatic two-loop computation of the lightest
Higgs-boson mass shows that, even for the extreme conditions analyzed
in~\cite{EN}, the numerical differences between both methods remain of
about 2.5~GeV~\cite{Heinpriv}.}.  In addition, $(\Delta
\Pi^S)^{\tilde{t},\tilde{b}} (s)$ is the one-loop scale-independent
contribution induced by the top and bottom squarks individually.  
Finally, $(\widetilde{\Pi}^S)^{tb} (s)$ represents the top- and
bottom-quark contributions to the CP-even Higgs-boson self-energies.

The same procedure may be used to obtain the RG-improved
analytic forms  of the remaining neutral Higgs-boson  self-energies:
$\widehat{\Pi}^{SP}(s)$  and $\widehat{\Pi}^{P}(s)$. Thus, in the weak
basis  $\{\phi_1,\phi_2,a_1,a_2\}$,    all   the   RG-improved neutral
Higgs-boson self-energies   may    be  summarized  by    the  $4\times
4$-dimensional matrix
\begin{equation}
  \label{RG0}
\widehat{\Pi}^N (s)\ =\ \left(\! \begin{array}{cc}
\widehat{\Pi}^S (s) & \widehat{\Pi}^{SP}(s) \\
\big(\widehat{\Pi}^{SP} (s)\big)^T & 
\widehat{\Pi}^P (s)\end{array}\!\right)\,.
\end{equation}
In the effective-potential limit $s\to  0$, the  would-be  Goldstone
bosons $G^\pm$ and $G^0$  defined in (\ref{rot}) decouple from the
charged and neutral Higgs-boson matrices.   As a consequence, the ranks
of the  respective self-energy  matrices given by (\ref{RGplus})
and (\ref{RG0}) get reduced by one unit.  To  make this last property
explicit, we express the RG-improved charged and  neutral
Higgs-boson self-energies $\widehat{\Pi}^\pm (s)$ and $\widehat{\Pi}^N
(s)$ in the weak bases  $\{ G^+,H^+\}$ and   $\{\phi_1, \phi_2, a,
G^0\}$, respectively.  In the following we assume that such a
change of  weak   bases  has  already been made using (\ref{rot}).

It is worth noticing that the RG-improved charged and neutral Higgs-boson
self-energies include, through the terms $\Delta\Pi^\pm$, $\Delta\Pi^S$,
etc., potentially large two-loop non-logarithmic contributions from
third-generation squarks, which are induced by their one-loop threshold
effects on the top- and bottom-quark Yukawa couplings $h_t$ and $h_b$.  
Since these effects were extensively studied in~\cite{CEPW} within the
MSSM with explicit radiative CP violation, we do not embark on this
topic here as well.  In the next section, instead, we calculate
analytically the pole masses of the charged and neutral Higgs bosons, and
discuss some field-theoretic subtleties relevant to this calculation.

\setcounter{equation}{0}
\section{Higgs-Boson Pole Masses}

We now concentrate on field-theoretic issues related to the proper
definition of the physical pole masses of the unstable charged and neutral
Higgs bosons.  In the MSSM with explicit radiative CP violation, as well
as being almost degenerate in mass, the two heaviest neutral Higgs bosons
may mix with one another very strongly.  This is a novel feature absent
from the CP-invariant MSSM~\cite{APLB}. In particular, a proper
description of the strong CP-violating mixing of the heavier Higgs bosons
requires a resummation approach that takes properly into account unstable
particle-mixing effects~\cite{APRL} in a way consistent with unitarity and
gauge invariance. For this, one might consider the resummation approach
implemented by the pinch technique (PT)~\cite{PT}. In general, when
naively resummed, off-shell Higgs-boson self-energies depend on the choice
of the gauge-fixing parameter, due to loops involving $W^\pm$ and $Z$
gauge bosons. However, this issue of gauge invariance does not directly
affect the present computation of the Higgs-boson pole masses, since we
neglect the subdominant contributions due to gauge-boson loops. Moreover,
it is known~\cite{Sirlin} that pole masses are by themselves
gauge-invariant quantities.

To start with, we consider the charged Higgs-boson  sector. The
dispersion  of the  charged  Higgs fields  is governed by  the inverse
propagator matrix
\begin{equation}
  \label{Iplus}
\hat{\Delta}^{-1}_\pm (s) \ =\  s {\bf 1}_2\: +\: \widehat{\Pi}^\pm (s)\,,
\end{equation}
where $\widehat{\Pi}^\pm (s)$ is the RG-improved self-energy matrix of
the charged Higgs fields expressed in  the weak basis $\{ G^\pm, H^\pm
\}$.  Note that  $\widehat{\Pi}^\pm (s)$ defined in (\ref{RGplus})
includes  the tree-level  contributions and so  does $\hat{\Delta}_\pm
(s)$.  Therefore, it is  more convenient to rewrite $\widehat{\Pi}^\pm
(s)$ as
\begin{equation}
  \label{Phatch}
\widehat{\Pi}^\pm (s)\ =\ \left(\! \begin{array}{cc}
\widehat{\Pi}_{G^+G^-} (s) & \widehat{\Pi}_{G^+H^-} (s) \\
\widehat{\Pi}_{H^+G^-} (s) & -\,M^{2(0)}_{H^+}\: +\:
\widehat{\Pi}_{H^+H^-} (s) \end{array} \!\right)\,,
\end{equation} 
where $M^{2(0)}_{H^+}$    is the  mass eigenvalue  of   the tree-level
charged Higgs-boson  mass  matrix ${\cal  M}^{2\,(0)}_{\pm}$  given by
(\ref{M0plus}). The gauge-independent   pole   masses  of    the
$G^+$-$H^+$ system   are determined by  the vanishing  of ${\rm det}\,
\hat{\Delta}^{-1}_\pm   (s)$.    Thus,   the   complex   pole   masses
$\hat{s}_{G^+}$ and $\hat{s}_{H^+}$ of  the $G^+$ and $H^+$ bosons are
found to be
\begin{eqnarray}
  \label{poles}
\hat{s}_{G^+ (H^+)} &=& \widehat{M}^2_{G^+(H^+)}\: -\: 
i\widehat{M}_{G^+(H^+)}\,\widehat{\Gamma}_{G^+(H^+)}\nonumber\\ 
&=& -\, \frac{1}{2}\, \bigg[\, {\rm Tr}\,
\widehat{\Pi}^\pm (\hat{s}_{G^+ (H^+)})\ -\,(+)\
\sqrt{{\rm Tr}^2\,
\widehat{\Pi}^\pm (\hat{s}_{G^+ (H^+)})\: -\: 4\,{\rm det}\,
\widehat{\Pi}^\pm (\hat{s}_{G^+ (H^+)})}\,\,\bigg]\, ,\qquad
\end{eqnarray}
where $\widehat{\Gamma}_{G^+(H^+)}$ correspond to the width of
the corresponding particles.
Employing the fact that ${\rm det}\, \widehat{\Pi}^\pm (0) = 0$ implies
$\hat{s}_{G^+}  = 0$, as follows from the Goldstone
theorem.  Since it   is practically impossible  to solve (\ref{poles}) 
analytically for $\hat{s}_{H^+}$,  we  shall determine  the  pole
mass  of the charged Higgs  boson $H^+$ using a perturbative expansion
with respect to the loop factors  $1/(16\pi^2)^n$, where $n$ indicates
the number of loops.  To this end, we consider the relation
\begin{equation}
  \label{sHplus}
\hat{s}_{H^+}\ =\ M^{2\,(0)}_{H^+}\: -\: 
\widehat{\Pi}_{H^+H^-} (\hat{s}_{H^+})\: +\:
\frac{\widehat{\Pi}_{H^+G^-} (\hat{s}_{H^+})\
\widehat{\Pi}_{G^+H^-}(\hat{s}_{H^+})}{  
\hat{s}_{H^+}\: +\:  \widehat{\Pi}_{G^+G^-}(\hat{s}_{H^+})}\ ,
\end{equation}
which  is completely equivalent to (\ref{poles}).  Within  the
framework of  perturbation  theory, a self-energy  function $\Pi (s)$,
e.g.\ $\widehat{\Pi}_{H^+H^-} (s)$, may be expanded loopwise as
\begin{equation}
  \label{Ploop}
\Pi (s) \ =\ \Pi^{(1)}(s)\: +\:  \Pi^{(2)}(s)\: +\: \cdots\: 
+\: \Pi^{(n)}(s)\: +\: \cdots .
\end{equation}
Obviously,  at  zeroth  order  we have   $\widehat{M}^{2\,(0)}_{H^+} =
M^{2\,(0)}_{H^+}$   and $\widehat{\Gamma}^{(0)}_{H^+} =   0$,  and at
one-loop order $\widehat{M}^{2\,(1)}_{H^+}   = 
\widehat{M}^{2\,(0)}_{H^+}\:
-\:            {\rm         Re}\,         \widehat{\Pi}^{(1)}_{H^+H^-}
(\widehat{M}^{2\,(1)}_{H^+})$       and       $\widehat{M}^{(1)}_{H^+}
\widehat{\Gamma}^{(1)}_{H^+} = {\rm Im}\, \widehat{\Pi}^{(1)}_{H^+H^-}
(\widehat{M}^{2\,(1)}_{H^+})$.     Making    use    of the    emerging
perturbative structure, we find at two-loop order
\begin{eqnarray}
  \label{Mch2}
\widehat{M}^{2\,(2)}_{H^+} &=& \widehat{M}^{2\,(0)}_{H^+}\: -\:  {\rm
  Re}\, \Big(\,\widehat{\Pi}^{(1)}_{H^+H^-} + 
\widehat{\Pi}^{(2)}_{H^+H^-}\Big)\: -\:
{\rm Im}\, \widehat{\Pi}^{(1)\prime}_{H^+H^-}\, 
{\rm Im}\, \widehat{\Pi}^{(1)}_{H^+H^-}\nonumber\\
&&+\, \frac{1}{\widehat{M}^{2\,(2)}_{H^+}}\,
{\rm Re}\, \Big(\, \widehat{\Pi}^{(1)}_{H^+G^-}\, 
\widehat{\Pi}^{(1)}_{G^+H^-}\,\Big)\,,\\
  \label{GMch2}
\widehat{M}^{(2)}_{H^+}\, \widehat{\Gamma}^{(2)}_{H^+} &=&
{\rm Im}\, \Big(\,\widehat{\Pi}^{(1)}_{H^+H^-} + 
\widehat{\Pi}^{(2)}_{H^+H^-}\,\Big)\: -\: {\rm Re}\,
\widehat{\Pi}^{(1)\prime}_{H^+H^-}\,  
{\rm Im}\, \widehat{\Pi}^{(1)}_{H^+H^-}\nonumber\\
&& -\, \frac{1}{\widehat{M}^{2\,(2)}_{H^+}}\,
{\rm Im}\, \Big(\, \widehat{\Pi}^{(1)}_{H^+G^-}\, 
\widehat{\Pi}^{(1)}_{G^+H^-}\,\Big)\,.
\end{eqnarray}
Here,  all self-energies involved  are understood  to be  evaluated at
$\widehat{M}^{2\,(2)}_{H^+}$.    Since   we   are  interested   in   a
perturbative one-loop computation of the charged Higgs-boson pole mass
that is  improved   by  including   two-loop   leading  logarithmic
corrections  and threshold effects,  we should  consistently disregard
higher-order   terms  occurring  in   $\widehat{M}^{(2)}_{H^+}$.
Specifically, since the third  and fourth terms on the right-hand side of 
(\ref{Mch2})
do  not receive  leading logarithmic  contributions, they may be
omitted~\footnote[2]{Moreover, we have verified in a number of 
cases that these higher-order 
terms indeed have negligible numerical impact on the neutral  Higgs-boson 
mass spectrum.}.

In the MSSM with explicit radiative CP  violation, it is convenient to
use as input the physical charged Higgs-boson mass $\widehat{M}_{H^+}$
instead  of ${\rm  Re}\,\bar{m}^2_{12}$  in (\ref{Mbarplus}).  The
free Lagrangian  parameter ${\rm Re}\,\bar{m}^2_{12}$  also enters the
neutral Higgs-boson  self-energy matrix $\widehat{\Pi}^N  (s)$ through
the two-loop Born-improved  mass  matrix $\overline{\cal M}^2_S$  
in (\ref{RGeven}) (and likewise $\overline{\cal  M}^2_P$).  We may 
therefore eliminate the quantity ${\rm  Re}\,\bar{m}^2_{12}$ appearing in  
$\widehat{\Pi}^N (s)$ in favour of $\widehat{M}^2_{H^+}$, using the relation
\begin{equation}
  \label{Rem12}
\frac{{\rm Re}\,\bar{m}^2_{12}}{\sin\beta\,\cos\beta}\ =\
\widehat{M}^2_{H^+}\: -\: \frac{1}{2}\, \bar{\lambda}_4\,
v^2\: +\: {\rm Re}\, \widehat{\Pi}_{H^+ H^-} (\widehat{M}^2_{H^+}) \, .
\end{equation}
In (\ref{Rem12}),  all  kinematic  parameters, with the exception of the
scale-independent  quantity $\widehat{M}^2_{H^+}$, are evaluated   at
$m_t$.  The  charged Higgs-boson  self-energy $\widehat{\Pi}_{H^+ H^-}
(s)$  is calculated  to  the one-loop order considered, and further
improved by including leading two-loop logarithms and threshold
effects.

The computation of the  pole masses of  the  neutral Higgs bosons  can
proceed along  similar lines.  As  in the charged Higgs-boson case, to
the  order of our perturbative  approximation, we  may neglect all the
self-energy  terms involving  the $G^0$   boson in  $\widehat{\Pi}^N$,
i.e.\  the terms  $\widehat{\Pi}_{G^0 G^0}$, $\widehat{\Pi}_{G^0  a}$,
$\widehat{\Pi}_{G^0      \phi_1}$  and  $\widehat{\Pi}_{G^0  \phi_2}$.
However,   we    cannot   neglect   the   other   matrix  elements  of
$\widehat{\Pi}^N$  in the  subspace  $\{  \phi_1,  \phi_2, a  \}$, and
especially those induced by   CP-violating loop effects.    

To better elucidate the above point, let us consider two Higgs scalars
$H$ and $A$ of opposite CP  quantum numbers: CP$(H) = +1$ and CP$(A) =
-1$.   For  instance,  for  $\tan\beta  \stackrel{>}{{}_\sim}  3$  and
$M_{H^+} \stackrel{>}{{}_\sim}  200$~GeV, one may identify $H$ 
approximately with
$\phi_1$ and  $A$ with $a$, the  two heaviest neutral  Higgs states in
the CP-invariant limit of the  theory. Further, the states $H$ and $A$
may  be considered  to be  nearly  degenerate, which  is exactly  what
happens in the  CP-invariant MSSM for the above  kinematic region of a
large $M_{H^+}$.  In the presence  of explicit sources of CP violation
(see  also  the  discussion  in  Section  2), $H$  and  $A$  will  mix
radiatively,  and one  then has  to  deal with  a strong  CP-violating
$H$-$A$ mixing  system~\cite{APLB,APRL}.  The dynamics  of the $H$-$A$
system  may   be  described   by  the  two-by-two   propagator  matrix
$\hat{\Delta}_n (s)$, whose inverse is given by
\begin{equation}
  \label{DHA}
\hat{\Delta}^{-1}_n   (s)\ =\ s{\bf 1}_2\: +\: \widehat{\Pi}^n (s)
\ =\ s{\bf 1}_2\: -\: \left(\!\begin{array}{cc}
M^{2\,(0)}_H\: -\: \widehat{\Pi}_{HH}(s) & -\, \widehat{\Pi}_{HA}(s)\\
-\, \widehat{\Pi}_{AH}(s) & M^{2\,(0)}_A\: -\: \widehat{\Pi}_{AA}(s)
\end{array}\!\right)\, .
\end{equation}
As before, the pole masses $\hat{s}_H$ and $\hat{s}_A$  of the $H$ and
$A$ bosons may be obtained as solutions of the characteristic equation
${\rm det}\, \hat{\Delta}^{-1}_n (s)   =  0$.  The last equation   can
equivalently be decomposed into a coupled system of complex equations:
\begin{eqnarray}
  \label{sH}
\hat{s}_H &=& M^{2\,(0)}_H\: -\: \widehat{\Pi}_{HH} (\hat{s}_{H})\: +\:
\frac{\widehat{\Pi}_{HA} (\hat{s}_H)\ \widehat{\Pi}_{AH}(\hat{s}_H)}{  
\hat{s}_H\: -\: M^{2\,(0)}_A\: +\:  \widehat{\Pi}_{AA}(\hat{s}_H)}\,,\\
  \label{sA}
\hat{s}_A &=& M^{2\,(0)}_A\: -\: \widehat{\Pi}_{AA} (\hat{s}_{A})\: +\:
\frac{\widehat{\Pi}_{AH} (\hat{s}_A)\ \widehat{\Pi}_{HA}(\hat{s}_A)}{  
\hat{s}_A\: -\: M^{2\,(0)}_H\: +\:  \widehat{\Pi}_{HH} (\hat{s}_A)}\,.
\end{eqnarray}
Within   the framework of perturbation   theory,  the above non-linear
system of equations could be solved iteratively, where $\hat{s}_H$ and
$\hat{s}_A$ on the left-hand sides of (\ref{sH}) and   (\ref{sA}) 
represent higher degrees of iteration. Such a procedure, however, does 
not lead to convergent solutions when
\begin{equation}
  \label{proc}
\big|\, M^{2\,(0)}_H\: -\: \widehat{\Pi}_{HH} (s)\ -\
M^{2\,(0)}_A\: +\: \widehat{\Pi}_{AA} (s)\,\big|\ 
\stackrel{<}{{}_\sim}\ 2\,\big|\widehat{\Pi}_{HA} (s)\big|\, ,
\end{equation}
at $s\approx \widehat{M}^2_H  \approx  \widehat{M}^2_A$.  Here, it  is
also interesting to remark that the inequality (\ref{proc}) reflects a
kinematic regime of resonantly-enhanced CP violation in Higgs-mediated
processes~\cite{APRL}.

To avoid the above  problem of convergence  in the computation of pole
masses, we follow a different two-step approach, which we apply
directly to  the original $4\times 4$ self-energy
matrix  $\widehat{\Pi}^N   (s)$.  The   first  step  consists in
diagonalizing $\Pi^N (s)$ in the effective-potential limit  $s = 0$ by
means of an orthogonal matrix~$O$:
\begin{equation}
  \label{proc0}
O^T\, \widehat{\Pi}^N (s)\, O\ =\ \overline{\Pi}^N (s)\, , 
\end{equation}
with
\begin{equation}
  \label{Pdiag}
-\,\overline{\Pi}^N (s=0)\ =\ {\rm diag}\, 
\Big(\, M^2_{H_1},\ M^2_{H_2},\ M^2_{H_3},\, 0\, \Big)\, .
\end{equation}
Clearly, the analytic forms  of $M_{H_{1,2,3}}$ coincide with the
RG-improved Higgs-boson masses computed in~\cite{CEPW}. In particular,
we can check that the numerical results obtained by the two approaches
are  identical, provided  the effective-potential  charged Higgs-boson
mass $M_{H^+}$ instead of the pole mass $\widehat{M}_{H^+}$ is used as
an input. In the second step, we determine the perturbative 
one-loop pole masses
$\widehat{M}^2_{H_{1,2,3}}$  of the  three  Higgs  bosons  through the
relation:
\begin{equation}
  \label{poleH}
\widehat{M}^2_{H_i}\ =\ -\, {\rm Re}\,
{\overline{\Pi}}^N_{ii} (M^2_{H_i})\,, 
\end{equation}
with $i=1,2,3$.  

Clearly,  the above expression  is an  excellent approximation  to the
exact masses in  cases in which the mass eigenvalues  are not close to
each  other.  For nearly  degenerate Higgs  bosons which  are strongly
mixed   by  CP   violation,   e.g.\  for   $|M^2_{H_2}  -   M^2_{H_3}|
\stackrel{<}{{}_\sim}     |\overline{\Pi}^N_{23}     (M^2_{H_2})     +
\overline{\Pi}^N_{23} (M^2_{H_3})|$,  our two-step computation  of the
Higgs-boson  pole masses  should  be further  improved.  Indeed,  this
situation  is  equivalent  to  the  one we  encountered  above  for  a
strongly-mixed  $H$-$A$ system (cf.~(\ref{proc})).   In this  case, we
first define an average  Higgs-boson mass, $M^2_{H_{23}} = \frac{1}2\,
(M^2_{H_2} +  M^2_{H_3})$, and then  find the mass eigenvalues  of the
two-by-two  submatrix  ${\overline{\Pi}}^N_{ij} (M^2_{H_{23}})$  (with
$i,j = 2,3$), where the $(H_2,H_3)$ subspace is described by a formula
analogous to (\ref{poles}). In this case, however, all expressions are
defined  at the  scale  $M^2_{H_{23}}$ and  therefore the  appropriate
solutions may be found in a straightforward way.

\setcounter{equation}{0}
\section{Numerical results}

For  our quantitative analysis~\cite{subroutine}, we  adopt the benchmark 
scenario
of maximal CP violation (CPX) introduced in~\cite{CPX}:
\begin{eqnarray}
  \label{CPX}
\widetilde{M}_Q \!&=&\! \widetilde{M}_t\ =\ 
\widetilde{M}_b\ =\ M_{\rm SUSY}\,,\qquad
\mu \ =\ 4 M_{\rm SUSY}\,,\nonumber\\
|A_t| \!&=&\!  |A_b|\ =\ 2M_{\rm SUSY}\,,\qquad
|m_{\tilde{g}}| \ =\ 1~{\rm TeV}\,.
\end{eqnarray}
According    to   our phase conventions,   the   parameters  $\mu$ and
$m^2_{12}$  are  taken to be  real,   whilst the soft-trilinear Yukawa
couplings  $A_{t,b}$  and gluino    mass $m_{\tilde{g}}$ are  complex.
Thus, CP violation in the  MSSM Higgs sector is predominantly mediated
by the CP-odd phase ${\rm arg}\,  (A_{t,b})$ at the one-loop level, with
${\rm arg}\,  (m_{\tilde{g}})$ entering at the two-loop level.   

We recall that the CP-violating parameters $A_t$ and $m_{\tilde{g}}$ may
also lead to observable two- and three-loop contributions to the electron
and neutron electric dipole moments (EDMs)~\cite{CKP,APino,DDLPD,AbelEDM}.
Since these higher-loop EDM effects get enhanced for large values of
$\tan\beta$, cancellations among the different one- and higher-loop EDM
terms may be needed~\cite{IN,CEPW}. A cancellation down to the required
10~\% level is generally not very difficult to arrange in the CPX
scenario, even for relatively large values of $\tan\beta \approx 20$.

As was already discussed in~\cite{CPX}, the large values of $|\mu|$
together with the comparatively low values of the charged Higgs-boson
mass that are used below in the analysis of the CPX scenario are not
feasible in minimal supergravity models with radiative electroweak
symmetry breaking.  In general, the realization of the CPX scenario
requires non-standard boundary conditions at the high-energy input
scale.  Nevertheless, these non-standard boundary conditions might be
obtained, for instance, within superstring-inspired models, in which
SUSY is broken by the VEV's of the auxiliary components of
the moduli fields.

As a first example, we consider a CPX scenario with $\tan\beta = 5$,
$\widehat{M}_{H^+} = 150$~GeV and $M_{\rm SUSY} = 0.5$~TeV.  In
Fig.~\ref{fig:pole1}(a), we show numerical values for the
effective-potential and pole masses of the two lightest Higgs bosons
$H_1$ and $H_2$ as functions of ${\rm arg}\, (A_t) = {\rm arg}\,
(A_b)$, for two different gluino phases: ${\rm arg}\, (m_{\tilde{g}})
= 0$ and $90^\circ$.  In detail, the numerical estimates of the
effective-potential masses $M_{H_1}$ and $M_{H_2}$ are indicated by
solid lines for ${\rm arg}\, (m_{\tilde{g}}) = 0$, and by dash-dotted
lines for ${\rm arg}\, (m_{\tilde{g}}) = 90^\circ$. The corresponding
numerical results for the pole masses $\widehat{M}_{H_1}$ and
$\widehat{M}_{H_2}$ are given by the dashed and dotted lines. As can
be seen from Fig.~\ref{fig:pole1}(a), the difference between effective
potential and pole masses becomes significant only for large values of
the stop mixing $|X_t| = |A_t - \mu^*/\tan\beta|$.  This difference
can be as large as 3~GeV, for ${\rm arg}\, (A_t) \approx \pm
90^\circ$.

One may observe that for ${\rm arg}\, (m_{\tilde{g}}) = 0^\circ$, the
numerical values for the $H_1$- and $H_2$-boson masses are
approximately symmetric around the vertical line ${\rm arg}\, (A_t) =
0$, whilst they exhibit a noticeable asymmetry if ${\rm arg}\,
(m_{\tilde{g}}) = 90^\circ$.  This asymmetry between positive and
negative values of ${\rm arg}\, (A_t)$ may be as large as 3~GeV. Its
origin can be attributed to the CP-violating stop threshold effects on
the top-quark Yukawa coupling. The main effect of these threshold
corrections is to change the absolute value of the top-quark Yukawa
coupling at the heavy squark-mass scale, which determines the value of
the scale-independent contributions to the effective potential. Taking
into account the expression for $h_t$ presented in~\cite{CEPW}, the
dominant contribution to these threshold corrections is proportional
to $A_t \, m_{\tilde{g}}$.

Since the tree-level top quark Yukawa coupling is real, for zero
gluino-mass phase, a change in sign of the phase of $A_t$ leads to a
change in sign of the imaginary part of the threshold correction,
without inducing any change in the absolute value of the top quark
Yukawa coupling at the one-loop level. For ${\rm arg}\,
(m_{\tilde{g}}) = 90^\circ$, instead, choosing ${\rm arg}\, (A_t) =
90^\circ \; (-90^\circ)$ leads to a threshold correction that reduces
(enhances) the absolute value of $h_t$. Indeed, the product of $A_t$
and $m_{\tilde{g}}$ reduces to a negative (positive) real number for
this particular choice of phases and therefore the situation is
similar to the CP-conserving case in which positive (negative) values
of $A_t$ enhance (reduce) $|h_t|$. As is clear from
Fig.~\ref{fig:pole1}(a), an increase (reduction) of the top quark
Yukawa couplings leads to a corresponding increase (reduction) of the
lightest Higgs-boson mass.  In this respect, it is interesting to
notice that the numerical values for ${\rm arg}\, (m_{\tilde{g}}) = -
90^\circ$, which were not plotted in Fig.~\ref{fig:pole1}(a), exhibit
an asymmetry as well.  In fact, the numerical lines for ${\rm arg}\,
(m_{\tilde{g}}) = - 90^\circ$ are almost mirror-symmetric with respect
to those for ${\rm arg}\, (m_{\tilde{g}}) = 90^\circ$ around the ${\rm
arg}\, (A_t) = 0$ line, in agreement with what is expected from the
argument presented above.

In Fig.~\ref{fig:pole1}(b),  we  present  numerical estimates  of  the
effective   Higgs-$Z$-$Z$ couplings  squared  $g^2_{H_iZZ}$ (with
$i=1,2,3$) versus ${\rm  arg}\,  (A_t) =  {\rm arg}\, (A_b)$,  for the
above  two choices of     the  gluino phase, i.e.~for  ${\rm    arg}\,
(m_{\tilde{g}})  = 0$ and $90^\circ$.  Note that $g_{H_iZZ}$ represent
the effective Higgs couplings to the $Z$ boson  normalized to their SM
value; the   precise determination  of  $g_{H_iZZ}$ in  terms   of the
orthogonal  Higgs-mixing    matrix    $O$      may   be    found    
in~\cite{PW,CEPW,CPX}.  {}From Fig.~\ref{fig:pole1}(b), we see that
the  predictions for $g^2_{H_iZZ}$ are affected in a minor way  by the  
value of the gluino phase. 

In Fig.~\ref{fig:pole2}, we show numerical estimates for the $H_1$-
and $H_2$-boson masses and for the effective Higgs couplings to the
$Z$ boson in a CPX scenario with $M_{\rm SUSY} = 1$~TeV.  Again, we
observe very similar features as in Fig.~\ref{fig:pole1}, namely the
difference between the effective-potential and pole masses may be as
large as 3~GeV, while there is no observable difference in the
effective Higgs-$Z$-$Z$ couplings.

The above results show that the differences between the pole masses and
those previously computed in the effective-potential approximation 
for light neutral Higgs bosons are small, of about the same size of
the possible uncertainties coming from higher-order loop corrections. 
The effects become more relevant in the case of a heavy and 
strongly-mixed Higgs sector, like that obtained in the CPX scenario
for relatively large values of the charged Higgs-boson mass.

In Fig.~\ref{fig:pole3}, we display numerical values for the masses of
the two heaviest neutral Higgs bosons, $H_2$ and $H_3$, as functions
of ${\rm arg}\, (A_t)$, in a CPX scenario with $\tan\beta = 5$,
$M_{\rm SUSY} = 0.5$~TeV and ${\rm arg}\, (m_{\tilde{g}}) =
90^\circ$. Going from the upper to the lower panel in
Fig.~\ref{fig:pole3}, we discretely vary the charged Higgs-boson pole
mass: $\widehat{M}_{H^+} = 200,\ 400,\ 600$~GeV.  Numerical results
pertaining to the effective-potential masses are indicated by solid
lines, while the results of pole masses are given by the dashed lines.

To  understand the  results presented  in Fig.~\ref{fig:pole3},  it is
important  to notice that  the lightest  stop mass  values in  the CPX
scenario for $\tan  \beta = 5$ and $M_{\rm SUSY} =  0.5$ TeV vary from
250 to 450 GeV, with values diminishing for larger values of the phase
of  $A_t$. Whilst  a light  Higgs boson  cannot decay  into a  pair of
supersymmetric  particles,   this  decay   channel  opens  up   for  a
sufficiently  heavy  charged  Higgs  boson.   An  enhancement  in  the
Higgs-boson  pole-mass shift  for  Higgs-boson masses  above the  stop
production  threshold  is  to  be  expected~\footnote{Since  $\mu$  is
relatively large,  chargino/neutralino on-shell effects  are, instead,
rather suppressed in the CPX scenario.}.  This effect is quite visible
in Fig.~\ref{fig:pole3}.  The pole masses start presenting significant
differences   from   those   computed   in   the   effective-potential
approximation, with the maximal  differences observed at the values of
the Higgs-boson masses equal to  the threshold for the production of a
pair   of  top   squarks.   This   threshold  is   quite   visible  in
Fig.~\ref{fig:pole3}, for which sizeable mass differences of about ten
percent of the  Higgs-boson mass are obtained. In  particular, this is
clearly shown in the cases of $\widehat{M}_{H^+} = 600$~GeV, for which
mass  shifts of about  50~GeV are  obtained~\footnote{ For  such large
values  of the charged  Higgs-boson mass,  the lightest  neutral Higgs
boson  $H_1$ has  SM-like properties,  due  to the  decoupling of  the
supersymmetric particles  and the heavy  Higgs bosons.}. We  note that
the threshold effect on the $H_3$ mass is much smaller than on that of
$H_2$, due  to a  significantly smaller coupling  of the $H_3$  to the
lighter stop quarks in the parameter range studied.

In this respect, it is instructive to compare Fig.~\ref{fig:pole3}
with Fig.~\ref{fig:pole4}.  Whilst in Fig.~\ref{fig:pole3} we
considered $M_{\rm SUSY} = $ 0.5 TeV, in Fig.~\ref{fig:pole4} we
considered $M_{\rm SUSY} =$ 1 TeV.  Contrary to the case of a light
Higgs sector, in the case of a heavy Higgs sector, the increase in the
SUSY-breaking scale has dramatic effects. The reason is that the
lightest stop particle becomes much heavier in this case, and
therefore the Higgs-boson masses are far away from the stop threshold.
The mass shift coming from the computation of the Higgs-boson pole
masses is never larger than about 3 \% in this case. In fact, this
result depends only slightly on the exact value of the heavy
Higgs-boson masses considered in Fig.~\ref{fig:pole4}.

It is important to stress that the shift in the pole masses tends to
increase the mass difference between the two neutral heavy Higgs
states. In the effective-potential approximation, this mass difference
is controlled by $|\mu A_t|$ (see~\cite{CMW}, for instance), becoming
smaller for larger values of the Higgs-boson masses (see the solid
lines in Figs.~\ref{fig:pole3} and \ref{fig:pole4}). For the reasons
explained above, the difference in pole masses between the two heavier
states may actually be larger than the effects on light states,
something that is quite apparent in Fig.~\ref{fig:pole3}. This
property may have important implications for physics at the LHC and at
the future high-energy electron-positron and muon colliders.

We  conclude this  section  with  a brief  discussion  of the  charged
Higgs-boson  pole mass~$\widehat{M}_{H^+}$.   As  in the  case of  the
neutral Higgs bosons, the effective-potential charged Higgs-boson mass
$M_{H^+}$  differs from  its pole  value  $\widehat{M}_{H^+}$.  Either
$M_{H^+}$  or $\widehat{M}_{H^+}$  could be  used equally  well  as an
input for the evaluation of the neutral Higgs-boson mass spectrum.  If
the same values for  $M_{H^+}$ and $\widehat{M}_{H^+}$ were considered
as inputs,  they would  result in slightly  different values  of ${\rm
Re}\,  \bar{m}^2_{12}$  and  so  in  modest  changes  in  the  neutral
Higgs-boson  masses.  Conversely,   the  mass  difference  $M_{H^+}  -
\widehat{M}_{H^+}$ can  be derived by finding values  of $M_{H^+}$ and
$\widehat{M}_{H^+}$  that give  the same  predictions for  the neutral
Higgs-boson  mass.   In  Fig.~\ref{fig:pole5},  we  present  numerical
estimates for this mass  difference $M_{H^+} - \widehat{M}_{H^+}$ as a
function of the charged Higgs-boson pole mass $\widehat{M}_{H^+}$, for
CPX  scenarios with  $\tan\beta =  5$, and  $M_{\rm SUSY}  =  0.5$ and
1~TeV.  In  these estimates, we  also select different phases  for the
CP-violating parameters $m_{\tilde{g}}$  and $A_t$: ${\rm arg}\, (A_t)
=  {\rm  arg}\,  (m_{\tilde{g}})  =  0$ (solid),  ${\rm  arg}\,  (A_t)
=90^\circ$  and  ${\rm arg}\,  (m_{\tilde{g}})  =  0$ (dashed),  ${\rm
arg}\, (A_t)  = {\rm arg}\, (m_{\tilde{g}}) =  90^\circ$ (dotted), and
${\rm  arg}\, (A_t)  =-90^\circ$  and ${\rm  arg}\, (m_{\tilde{g}})  =
90^\circ$ (dash-dotted).   We observe that  the shifts in  the charged
Higgs-boson mass are  larger in the CPX scenario  with $M_{\rm SUSY} =
0.5$~TeV  shown in  Fig.~\ref{fig:pole5}(a) than  in the  CPX scenario
with $M_{\rm  SUSY} = 1$~TeV shown in  Fig.~\ref{fig:pole5}(b), due to
the  fact  that stop-sbottom  production  threshold  effects are  more
significant in the first scenario.

\setcounter{equation}{0}
\section{Conclusions} 

In this article we have computed the Higgs-boson pole masses within
the minimal supersymmetric standard model with explicit radiative CP
violation in the Higgs sector~\cite{subroutine}.  One-loop self-energy
effects included in the definition of the pole masses are computed
diagrammatically, while higher-order effects are computed via a RG
improvement similar to that used previously in the effective-potential
approximation.

In general, the effect on the Higgs-boson masses depends on the value
of the supersymmetric particle masses: the closer the Higgs-boson mass
is to the production threshold for a pair of supersymmetric particles,
the larger becomes the difference between the pole masses and that
computed in the effective-potential approach. Therefore, the relative
shifts between pole and running masses are constrained to be much
smaller in a light Higgs sector, where the masses are below the
production threshold for top squarks.  Still, for the $H_1$-boson
sector, the absolute mass-difference value $|M_{H_1} -
\widehat{M}_{H_1}|$ can be as large as 3~GeV. There is also a
significant asymmetry between positive and negative values of ${\rm
arg}\, (A_t)$, when ${\rm arg}\, (m_{\tilde{g}}) $ is large, e.g.\
${\rm arg}\, (m_{\tilde{g}}) = \pm 90^\circ$, which is mainly
associated with the supersymmetric threshold corrections to the
top-quark Yukawa coupling.

The effects of computing the pole masses versus the running masses in
the heavier Higgs sector may, instead, be very significant,
particularly when the $H_2$-boson mass is close to the stop production
threshold. We find shifts in masses that can be as large as 10 \% of
the physical Higgs-boson masses. In addition, these mass shifts tend
to increase the mass splitting between the two neutral heavy Higgs
bosons, and therefore may have important implications for Higgs
physics at the LHC, or at future Higgs factories such as linear
electron-positron and muon colliders.

\subsection*{Acknowledgements}
The work of C.W. is supported in part by the US DOE,  Division of 
High-Energy Physics, under the contract no.: W-31-109-ENG-38.

\newpage

\def\theequation{\Alph{section}.\arabic{equation}}
\begin{appendix}
\setcounter{equation}{0}
\section{Higgs-Boson Couplings to Squarks}

We present in this Appendix the couplings of the  Higgs  bosons to 
squarks, incorporating CP violation. 

The task of computing the Higgs-boson self-energies simplifies
considerably if the Higgs-boson couplings are expressed in terms of
$2\times 2$ matrices in the weak basis of left-handed and right-handed
squarks, $\tilde{q}_L$ and $\tilde{q}_R$. Before listing the Higgs-boson
couplings, it is useful to recall some of the standard relations
associated with the squark sector of the third generation. The stop and
sbottom mass matrices may conveniently be written as
\begin{equation}
  \label{Mscalar}
\widetilde{\cal M}^2_q \ =\ \left( \begin{array}{cc}
\tilde{M}^2_Q\, +\, m^2_q\, +\, \cos 2\beta M^2_Z\, ( T^q_z\, -\,
Q_q \sin^2\theta_w ) & m_q (A^*_q - \mu R_q )\\ 
m_q (A_q - \mu^* R_q) & \hspace{-0.2cm}
\tilde{M}^2_q\, +\, m^2_q\, +\, \cos 2\beta M^2_Z\, Q_q \sin^2\theta_w 
\end{array}\right)\, ,
\end{equation}
with $q=t,b$, $Q_t = 2/3$, $Q_b = -1/3$, $T^t_z = - T^b_z = 1/2$, $R_b
= \tan\beta = v_2/v_1$, $R_t = \cot\beta$, $M^2_Z = \frac{1}4 (g^2_w +
g'^2) v^2$ and     $\sin^2\theta_w   = g'^2/(g^2_w +    g'^2)$.     In
(\ref{Mscalar}),    $\tilde{M}^2_Q$    and   $\tilde{M}^2_q$   are
soft SUSY-breaking   masses  for  the left-handed   and   right-handed
third-generation squarks.  As usual,  $\widetilde{\cal M}^2_q$ may  be
diagonalized by a unitary transformation $U^q$. The unitary matrix
$U^q$ relates the   weak  ($\tilde{q}_L,\ \tilde{q}_R$) and mass
eigenstates ($\tilde{q}_1,\ \tilde{q}_2$) through:
\begin{equation}
  \label{Rscalar}
\left( \begin{array}{c} \tilde{q}_L \\ \tilde{q}_R \end{array} \right)
\ =\ U^q\,
\left( \begin{array}{c} \tilde{q}_1 \\ \tilde{q}_2
  \end{array}\right)\, ;\quad 
U^q \ =\ 
\left( \begin{array}{cc} 1 & 0 \\ 0 & e^{i\delta_q} \end{array} \right)
\left( \begin{array}{cc} \cos\theta_q & \sin\theta_q \\
          -\sin\theta_q & \cos\theta_q \end{array} \right)\, ,
\end{equation}
where $\delta_q = {\rm  arg} (A_q -  R_q \mu^*)$ and the mixing angles
$\theta_q$ can uniquely be determined by
\begin{eqnarray}
  \label{theta}
\cos\theta_q \!&=&\! \frac{m_q |A_q - R_q \mu^*|}{ 
\sqrt{m^2_q |A_q - R_q \mu^*|^2\, +\, [ (\widetilde{\cal M}^2_q)_{11}
-  m^2_{\tilde{q}_1} ]^2} }\ ,\nonumber\\
\sin\theta_q \!&=&\! \frac{ |(\widetilde{\cal M}^2_q)_{11}
-  m^2_{\tilde{q}_1}|}{ 
\sqrt{m^2_q |A_q - R_q \mu^*|^2\, +\, [ (\widetilde{\cal M}^2_q)_{11}
-  m^2_{\tilde{q}_1} ]^2} }\ .
\end{eqnarray}
Furthermore,  the  mass eigenvalues  of  $\widetilde{\cal M}^2_q$  are
easily found to be
\begin{eqnarray}
  \label{Mq12}
m^2_{\tilde{q}_1 (\tilde{q}_2)} \!\!&=&\!\! \frac{1}{2}\ \bigg\{
\tilde{M}^2_Q + \tilde{M}^2_q + 2m^2_q + T^q_z \cos 2\beta M^2_Z \nonumber\\
\!\!&&\!\! +(-)\ \sqrt{ \Big[ \tilde{M}^2_Q - \tilde{M}^2_q 
+ \cos 2\beta M^2_Z ( T^q_z - 2Q_q \sin^2\theta_w )\, \Big]^2\, 
                   +\, 4m^2_q |A^*_q -\mu R_q |^2 }\ \bigg\}.\qquad
\end{eqnarray}

In the following, we list individually the Higgs-boson couplings
to top and bottom  squarks, employing a $2\times  2$-dimensional matrix
representation:
\begin{eqnarray}
  \label{phisq}
\Gamma^{\phi_1 \tilde{t}^* \tilde{t}} & =& \left( \begin{array}{cc}
-\,\frac{1}{4}\,(g^2_w - \frac{1}{3} g'^2)\,v_1 & 
                 \frac{1}{\sqrt{2}}\,h_t^*\mu\\
\frac{1}{\sqrt{2}}\,h_t\mu^* & -\,\frac{1}{3}\, g'^2\, v_1 
\end{array}\right)\,,\nonumber\\
\Gamma^{\phi_2 \tilde{t}^* \tilde{t}} & =& \left( \begin{array}{cc}
-\,|h_t|^2\,v_2\: +\: \frac{1}{4}\,( g^2_w - \frac{1}{3} g'^2)\, v_2 & 
-\,\frac{1}{\sqrt{2}}\,h^*_t A^*_t\\
-\,\frac{1}{\sqrt{2}}\,h_t A_t & -\,|h_t|^2\,v_2\: +\: \frac{1}{3}\,g'^2\, v_2 
\end{array}\right)\,,\nonumber\\
\Gamma^{\phi_1 \tilde{b}^* \tilde{b}} & =& \left( \begin{array}{cc}
-\,|h_b|^2\,v_1\: +\: \frac{1}{4}\,( g^2_w + \frac{1}{3} g'^2)\, v_1 & 
-\,\frac{1}{\sqrt{2}}\,h^*_b A^*_b\\
-\,\frac{1}{\sqrt{2}}\,h_b A_b & -\,|h_b|^2\,v_1\: +\: \frac{1}{6}\,g'^2\, v_1 
\end{array}\right)\,,\nonumber\\
\Gamma^{\phi_2 \tilde{b}^* \tilde{b}} & =& \left( \begin{array}{cc}
-\,\frac{1}{4}\,(g^2_w + \frac{1}{3} g'^2)\,v_2 & \frac{1}{\sqrt{2}}\,
h_b^*\mu\\
\frac{1}{\sqrt{2}}\,h_b\mu^* & -\,\frac{1}{6}\, g'^2\, v_2 
\end{array}\right)\,,\nonumber\\
  \label{asq}
\Gamma^{a_1 \tilde{t}^* \tilde{t}} & =& \frac{i}{\sqrt{2}}\,
\left( \begin{array}{cc} 0 & -\,h^*_t\mu \\ h_t\mu^* & 0 \end{array}\right)\,,
\qquad\ \
\Gamma^{a_2 \tilde{t}^* \tilde{t}} \ =\ \frac{i}{\sqrt{2}}\,
\left( \begin{array}{cc} 0 & h^*_tA_t^* \\ -\,h_tA_t & 0 \end{array}\right)\, ,
\nonumber\\
\Gamma^{a_1 \tilde{b}^* \tilde{b}} & =& \frac{i}{\sqrt{2}}\,
\left( \begin{array}{cc} 0 & -\,h^*_bA_b^* \\ 
h_b A_b & 0\end{array}\right)\,,\qquad
\Gamma^{a_2 \tilde{b}^* \tilde{b}} \ =\ \frac{i}{\sqrt{2}}\,
\left( \begin{array}{cc} 0 & h_b^*\mu \\ -\,h_b\mu^* & 0 \end{array}\right)\,,
\nonumber\\
  \label{plussq}
\Gamma^{\phi^+_1 \tilde{t}^* \tilde{b}} & =& 
\big( \Gamma^{\phi^-_1 \tilde{b}^* \tilde{t}}\,\big)^\dagger\ =\ 
\left( \begin{array}{cc}
-\,\frac{1}{\sqrt{2}}\,(|h_b|^2 - \frac{1}{2} g_w^2)\,v_1 & -\,h^*_b A^*_b\\
-\,h_t\mu^* & -\,\frac{1}{\sqrt{2}}\, h_t h^*_b\, v_2 
\end{array}\right)\,,\nonumber\\
\Gamma^{\phi^+_2 \tilde{t}^* \tilde{b}} & =& 
\big( \Gamma^{\phi^-_2 \tilde{b}^* \tilde{t}}\,\big)^\dagger\ =\ 
\left( \begin{array}{cc}
\frac{1}{\sqrt{2}}\,(|h_t|^2 - \frac{1}{2} g_w^2)\,v_2 & h_b^* \mu\\
h_tA_t & \frac{1}{\sqrt{2}}\, h_t h^*_b\, v_1 \end{array}\right)\, .
\end{eqnarray}
Without loss of generality, the  tree-level Yukawa couplings $h_t$ and
$h_b$ can be taken to be real and positive~\cite{CEPW}, i.e.\ the $t$-
and $b$-quark masses are given by  $m_t = \frac{1}{\sqrt{2}}\,h_t v_2$
and $m_b = \frac{1}{\sqrt{2}}\,h_b v_1$ at the tree level. Then, it is
not difficult to verify that
\begin{eqnarray}
\widetilde{\cal M}^2_t & =& {\rm diag}\, \big( \tilde{M}^2_Q,\,
\tilde{M}^2_t\big)\: -\: v_1\, \Gamma^{\phi_1 \tilde{t}^*\tilde{t}}
\: -\: v_2\, \Gamma^{\phi_2 \tilde{t}^* \tilde{t}}\,,\nonumber\\
\widetilde{\cal M}^2_b &=& {\rm diag}\, \big( \tilde{M}^2_Q,\,
\tilde{M}^2_b\big)\: -\: v_1\, \Gamma^{\phi_1 \tilde{b}^*\tilde{b}}
\: -\: v_2\, \Gamma^{\phi_2 \tilde{b}^* \tilde{b}}\,.
\end{eqnarray}

As well as trilinear couplings, Higgs  bosons  also have quadrilinear
couplings to squarks. These additional couplings are given by
\begin{eqnarray}
  \label{neutral4}
\Gamma^{\phi_1\phi_1 \tilde{t}^* \tilde{t}} & =& 
\Gamma^{a_1a_1 \tilde{t}^* \tilde{t}}\ =\ 
\left( \begin{array}{cc}
-\,\frac{1}{4}\,(g^2_w - \frac{1}{3} g'^2) &  0\\
0 & -\,\frac{1}{3}\, g'^2 \end{array}\right)\,,\nonumber\\
\Gamma^{\phi_2\phi_2 \tilde{t}^* \tilde{t}} & =& 
\Gamma^{a_2 a_2\tilde{t}^* \tilde{t}}\ =\ 
\left( \begin{array}{cc}
-\,|h_t|^2\: +\: \frac{1}{4}\,( g^2_w - \frac{1}{3} g'^2) & 0\\
0 & -\,|h_t|^2\: +\: \frac{1}{3}\,g'^2 \end{array}\right)\,,\nonumber\\
\Gamma^{\phi_1 \phi_1\tilde{b}^* \tilde{b}} & =& 
\Gamma^{a_1 a_1\tilde{b}^* \tilde{b}} \ =\ 
\left( \begin{array}{cc}
-\,|h_b|^2\: +\: \frac{1}{4}\,( g^2_w + \frac{1}{3} g'^2) & 0\\
0 & -\,|h_b|^2\: +\: \frac{1}{6}\,g'^2 \end{array}\right)\,,\nonumber\\
\Gamma^{\phi_2\phi_2\tilde{b}^* \tilde{b}} & =&
\Gamma^{a_2 a_2\tilde{b}^* \tilde{b}}\ =\ 
 \left( \begin{array}{cc}
-\,\frac{1}{4}\,(g^2_w + \frac{1}{3} g'^2) & 0 \\
0 & -\,\frac{1}{6}\, g'^2 \end{array}\right)\,,\nonumber\\
  \label{charge4}
\Gamma^{\phi^+_1\phi^-_1 \tilde{t}^* \tilde{t}} & =& 
\left( \begin{array}{cc}
-\,|h_b|^2\: +\: \frac{1}{4}\,(g^2_w + \frac{1}{3} g'^2) &  0\\
0 & -\,\frac{1}{3}\, g'^2 \end{array}\right)\,,\nonumber\\
\Gamma^{\phi^+_2\phi^-_2 \tilde{t}^* \tilde{t}} & =& 
\left( \begin{array}{cc}
-\, \frac{1}{4}\,(g^2_w + \frac{1}{3} g'^2) &  0\\
0 & -\, |h_t|^2\: +\: \frac{1}{3}\, g'^2 \end{array}\right)\,,\nonumber\\
\Gamma^{\phi^+_1\phi^-_1 \tilde{b}^* \tilde{b}} & =& 
\left( \begin{array}{cc}
-\, \frac{1}{4}\,(g^2_w - \frac{1}{3} g'^2) &  0\\
0 & -\,|h_b|^2\: +\: \frac{1}{6}\, g'^2 \end{array}\right)\,,\nonumber\\
\Gamma^{\phi^+_2\phi^-_2 \tilde{b}^* \tilde{b}} & =& 
\left( \begin{array}{cc}
-\, |h_t|^2\: +\: \frac{1}{4}\,(g^2_w - \frac{1}{3} g'^2) &  0\\
0 & -\, \frac{1}{6}\, g'^2 \end{array}\right)\, .
\end{eqnarray}

\setcounter{equation}{0}
\section{Higgs-Boson Self-Energies}

In this Appendix, we derive analytic expressions for the neutral
and charged   Higgs-boson  self-energies. We first  consider the
contributions of the third-generation squarks $\tilde{q} = \tilde{t},\ 
\tilde{b}$ to the  CP-even Higgs-boson self-energy transitions $\phi_j
\to \phi_i$.   Then,  the CP-even  Higgs-boson self-energies  shown in
Fig.~\ref{f1} may be computed by
\begin{eqnarray}
  \label{PiS}
\Pi^{S,(a)}_{ij} (s) & = & N_c\, \sum_{\tilde{q} =
\tilde{t},\tilde{b}}\, \int\!\frac{d^D k}{(2\pi)^D\, i}\ {\rm Tr}\,
\Big[\, i\Delta^{\tilde{q}} (k+p)\, i\Gamma^{\phi_i
\tilde{q}^*\tilde{q}}\, i\Delta^{\tilde{q}} (k)\, i\Gamma^{\phi_j
\tilde{q}^*\tilde{q}}\,\Big]\ ,\nonumber\\
\Pi^{S,(b)}_{ij} (s) & = & \delta_{ij}\, N_c\, \sum_{\tilde{q} =
\tilde{t},\tilde{b}}\, \int\!\frac{d^D k}{(2\pi)^D\, i}\ {\rm Tr}\,
\Big[\, i\Delta^{\tilde{q}} (k)\, i\Gamma^{\phi_i \phi_i
\tilde{q}^*\tilde{q}}\,\Big]\ ,
\end{eqnarray}
where $D = 4 - 2 \varepsilon$, $N_c = 3$ is the colour factor, and
\begin{equation}
  \label{Dsq}
\Delta^{\tilde{q}} (k)\ =\    U^q\  {\rm diag}\,  \big[   \big( k^2  -
m^2_{\tilde{q}_1}    \big)^{-1},\   \big(    k^2  -  m^2_{\tilde{q}_2}
\big)^{-1}\, \big]\  U^{q\dagger}
\end{equation}
is the  squark-propagator matrix written  in  the  weak  basis.  The
unitary  mixing  matrix  $U^q$  that  relates   the weak and mass
eigenstates   as   well  as     the $2\times   2$    coupling matrices
$\Gamma^{\phi_i \tilde{q}^*  \tilde{q}}$    and $\Gamma^{\phi_i \phi_i
\tilde{q}^* \tilde{q}}$ were already presented in Appendix A.

To evaluate the  self-energies  in  (\ref{PiS}), it is more
convenient to decompose the squark propagator as
\begin{equation}
  \label{Propsq}
\Delta^{\tilde{q}}(k)\ =\ \Delta^{\tilde{q}}_+ (k)\, {\bf 1}_2\ +\
\Delta^{\tilde{q}}_- (k)\, U^q\,\tau_3\, U^{q\dagger}\, ,
\end{equation}
with ${\bf 1}_2 = {\rm diag}\, (1, 1)$, $\tau_3 = {\rm diag}\, (1,-1)$
and
\begin{equation}
  \label{Ddec}
\Delta^{\tilde{q}}_\pm (k)\ =\ \frac{1}{2}\, \bigg(\, \frac{1}{k^2 -
m^2_{\tilde{q}_1} }\ \pm\ \frac{1}{k^2 - m^2_{\tilde{q}_2} }\,
\bigg)\, .
\end{equation}
With the above decomposition   of the squark propagators, the  neutral
Higgs-boson self-energies are analytically determined by
\begin{eqnarray}
  \label{PiSa}
\Pi^{S,(a)}_{ij} (s) & =& N_c\, \frac{1}{64\pi^2}\, \sum_{\tilde{q} =
\tilde{t},\tilde{b}}\,
\bigg\{\, \Big[\, B_0 (s,m^2_{\tilde{q}_1},m^2_{\tilde{q}_1})\: +\:
B_0 (s,m^2_{\tilde{q}_2},m^2_{\tilde{q}_2})\nonumber\\
&& +\, 2B_0 (s,m^2_{\tilde{q}_1},m^2_{\tilde{q}_2})\,\Big]\, 
{\rm Tr}\, \Big( \Gamma^{\phi_i \tilde{q}^*\tilde{q}}\,
\Gamma^{\phi_j \tilde{q}^*\tilde{q}}\Big)\nonumber\\
&&+\, \Big[\, B_0 (s,m^2_{\tilde{q}_1},m^2_{\tilde{q}_1})\: - \:
B_0 (s,m^2_{\tilde{q}_2},m^2_{\tilde{q}_2})\, \Big]\, 
\Big[\, {\rm Tr}\, \Big( \Gamma^{\phi_i \tilde{q}^*\tilde{q}}\,
\,U^q \tau_3 U^{q\dagger}\,\Gamma^{\phi_j \tilde{q}^*\tilde{q}}\Big)\nonumber\\
&&+\, {\rm Tr}\, \Big( \Gamma^{\phi_i \tilde{q}^*\tilde{q}}\,
\Gamma^{\phi_j \tilde{q}^*\tilde{q}}\,U^q \tau_3 U^{q\dagger}\Big)\,
\Big]\: +\: \Big[\, B_0 (s,m^2_{\tilde{q}_1},m^2_{\tilde{q}_1})\: +\: 
B_0 (s,m^2_{\tilde{q}_2},m^2_{\tilde{q}_2})\nonumber\\
&&-\, 2B_0 (s,m^2_{\tilde{q}_1},m^2_{\tilde{q}_2})\,\Big]\, 
{\rm Tr}\, \Big( \Gamma^{\phi_i \tilde{q}^*\tilde{q}}\,
U^q \tau_3 U^{q\dagger}\, \Gamma^{\phi_j \tilde{q}^*\tilde{q}}\,
U^q \tau_3 U^{q\dagger} \Big)\, \bigg\}\, ,\\
  \label{PiSb}
\Pi^{S,(b)}_{ij} (s) & =& -\,\delta_{ij}\, 
N_c\, \frac{1}{32\pi^2}\, \sum_{\tilde{q} =
\tilde{t},\tilde{b}}\,
\bigg\{\, \Big[\, A_0 (m^2_{\tilde{q}_1})\: +\: A_0
(m^2_{\tilde{q}_2})\, \Big]\, {\rm Tr}\, 
\Big( \Gamma^{\phi_i\phi_i \tilde{q}^*\tilde{q}}\Big)\nonumber\\
&&+\, \Big[\, A_0 (m^2_{\tilde{q}_1})\: -\: A_0
(m^2_{\tilde{q}_2})\, \Big]\, 
{\rm Tr}\, \Big( \Gamma^{\phi_i\phi_i \tilde{q}^*\tilde{q}}\, 
U^q \tau_3 U^{q\dagger}\Big)\, \bigg\}\, .
\end{eqnarray}
Here,   $B_0   (s,m^2_1,m^2_2)$ and  $A_0     (m^2)$   are the   usual
Pasarino--Veltman one-loop functions~\cite{PV}:
\begin{eqnarray}
  \label{B0} 
B_0 (s,m^2_1,m^2_2) &=& C_{\rm UV}\, -\, 
                     \ln \bigg(\frac{m_1m_2}{\mu^2}\bigg)\,
+\, 2\, +\, \frac{1}{s}\, \bigg[\, (m^2_2-m^2_1)\,
\ln\bigg(\frac{m_1}{m_2}\bigg)\nonumber\\ 
&&+\, \lambda^{1/2}(s,m^2_1,m^2_2)\,\, {\rm cosh}^{-1} \bigg(
\frac{m^2_1+m^2_2-s}{2m_1m_2}\bigg)\, \bigg]\, ,\\
A_0 (m^2) &=& m^2\, \Big[\,1\: +\: B_0  (0,m^2,m^2)\,\Big]\, ,
\end{eqnarray}
with  $C_{\rm UV} = \frac{1}{\varepsilon}  - \gamma_E +  \ln 4\pi$ and
$\lambda  (x,y,z)    = (x  -  y -     z)^2 - 4yz$. In  particular,
$B_0(s,m^2_1,m^2_2)$ simplifies further when evaluated at $s = 0$, to
\begin{equation}
  \label{B00}
B_0 (0, m^2_1, m^2_2)\ =\  C_{\rm UV}\:  -\: 
                     \ln \bigg(\frac{m_1m_2}{\mu^2}\bigg)\: +\: 1\: +\:
\frac{m^2_1 + m^2_2}{m^2_1 - m^2_2}\, \ln\bigg(\frac{m_2}{m_1}\bigg)\, .
\end{equation}
Moreover, $U^q \tau_3 U^{q\dagger}$ may be expressed entirely in terms
of the elements of the squark mass matrix $\widetilde{\cal M}^2_q$ and
of its squared mass-eigenvalues $m^2_{\tilde{q}_{1,2}}$, as
\begin{equation}
  \label{Utau}
U^q \tau_3 U^{q\dagger}\ =\ \frac{1}{m^2_{\tilde{q}_1} -
m^2_{\tilde{q}_2}}\, \left(\begin{array}{cc}
(\widetilde{\cal M}^2_q)_{11} - (\widetilde{\cal M}^2_q)_{22} & 
2\, (\widetilde{\cal M}^2_q)_{12} \\
2\, (\widetilde{\cal M}^{2*}_q)_{12} & 
(\widetilde{\cal M}^2_q)_{22} - (\widetilde{\cal M}^2_q)_{11}
\end{array} \right)\, .
\end{equation}
The  analytic  expressions  of  the CP-odd  Higgs-boson  self-energies
$\Pi^P_{ij}  (s)$ may  easily be  recovered from  the CP-even  ones by
replacing   the    coupling   matrices   $\Gamma^{\phi_i   \tilde{q}^*
\tilde{q}}$  and $\Gamma^{\phi_i  \phi_i \tilde{q}^*  \tilde{q}}$ with
$\Gamma^{a_i \tilde{q}^* \tilde{q}}$  and $\Gamma^{a_i a_i \tilde{q}^*
\tilde{q}}$ in (\ref{PiSa}) and (\ref{PiSb}).

The CP-violating scalar-pseudoscalar  transitions $\phi_j \to a_i$ may
also be calculated in a very analogous manner. Specifically, we find
\begin{eqnarray}
  \label{PiSP}
\Pi^{SP,(a)}_{ij} (s) & =& N_c\, \frac{1}{64\pi^2}\, \sum_{\tilde{q} =
\tilde{t},\tilde{b}}\,
\bigg\{\, \Big[\, B_0 (s,m^2_{\tilde{q}_1},m^2_{\tilde{q}_1})\: +\:
B_0 (s,m^2_{\tilde{q}_2},m^2_{\tilde{q}_2})\nonumber\\ 
&&+\, 2B_0 (s,m^2_{\tilde{q}_1},m^2_{\tilde{q}_2})\,\Big]\, 
{\rm Tr}\, \Big( \Gamma^{a_i \tilde{q}^*\tilde{q}}\,
\Gamma^{\phi_j \tilde{q}^*\tilde{q}}\Big)\nonumber\\
&&+\,\Big[\, B_0 (s,m^2_{\tilde{q}_1},m^2_{\tilde{q}_1})\: - \:
B_0 (s,m^2_{\tilde{q}_2},m^2_{\tilde{q}_2})\, \Big]\, \Big[\,
{\rm Tr}\, \Big( \Gamma^{a_i \tilde{q}^*\tilde{q}}\,
\,U^q \tau_3U^{q\dagger}\,\Gamma^{\phi_j\tilde{q}^*\tilde{q}}\Big)\nonumber\\
&&+\, {\rm Tr}\, \Big( \Gamma^{a_i \tilde{q}^*\tilde{q}}\,
\Gamma^{\phi_j \tilde{q}^*\tilde{q}}\,U^q \tau_3
U^{q\dagger}\Big)\,\Big]\: +\: 
\Big[\, B_0 (s,m^2_{\tilde{q}_1},m^2_{\tilde{q}_1})\: +\:
B_0 (s,m^2_{\tilde{q}_2},m^2_{\tilde{q}_2})\nonumber\\
&& -\,2B_0 (s,m^2_{\tilde{q}_1},m^2_{\tilde{q}_2})\,\Big]\, 
{\rm Tr}\, \Big( \Gamma^{a_i \tilde{q}^*\tilde{q}}\,
U^q \tau_3 U^{q\dagger}\, \Gamma^{\phi_j \tilde{q}^*\tilde{q}}\,
U^q \tau_3 U^{q\dagger} \Big)\, \bigg\}\, .
\end{eqnarray}
We  have checked that (\ref{PiSP}) is in  agreement with the
original derivation presented in~\cite{APLB}.

Finally,  the   analytic   expressions of    the charged   Higgs-boson
self-energy  transitions $\phi^-_j \to \phi^-_i$  take on an analogous
form
\begin{eqnarray}
  \label{Picha}
\Pi^{\pm,(a)}_{ij} (s) & =& N_c\, \frac{1}{64\pi^2}\, 
\bigg\{\, \Big[\, B_0 (s,m^2_{\tilde{t}_1},m^2_{\tilde{b}_1})\: +\:
B_0 (s,m^2_{\tilde{t}_1},m^2_{\tilde{b}_2})\: +\:
B_0 (s,m^2_{\tilde{t}_2},m^2_{\tilde{b}_1}) \nonumber\\
&&+\, B_0 (s,m^2_{\tilde{t}_2},m^2_{\tilde{b}_2})\,\Big]\, 
{\rm Tr}\, \Big( \Gamma^{\phi^+_i \tilde{t}^*\tilde{b}}\,
\Gamma^{\phi^-_j \tilde{b}^*\tilde{t}}\Big)\nonumber\\
&&+\, \Big[\, 
B_0 (s,m^2_{\tilde{t}_1},m^2_{\tilde{b}_1})\: -\:
B_0 (s,m^2_{\tilde{t}_1},m^2_{\tilde{b}_2})\: +\:
B_0 (s,m^2_{\tilde{t}_2},m^2_{\tilde{b}_1})\: -\: 
B_0 (s,m^2_{\tilde{t}_2},m^2_{\tilde{b}_2})\, \Big]\nonumber\\
&&\times\,{\rm Tr}\, \Big( \Gamma^{\phi^+_i \tilde{t}^*\tilde{b}}\,
\,U^b \tau_3 U^{b\dagger}\,\Gamma^{\phi^-_j \tilde{b}^*\tilde{t}}\Big)
 \nonumber\\
&&+\, \Big[\, 
B_0 (s,m^2_{\tilde{t}_1},m^2_{\tilde{b}_1})\: +\:
B_0 (s,m^2_{\tilde{t}_1},m^2_{\tilde{b}_2})\: -\:
B_0 (s,m^2_{\tilde{t}_2},m^2_{\tilde{b}_1})\: - \:
B_0 (s,m^2_{\tilde{t}_2},m^2_{\tilde{b}_2})\, \Big]\nonumber\\
&&\times\, {\rm Tr}\, \Big( \Gamma^{\phi^+_i \tilde{t}^*\tilde{b}}\,
\Gamma^{\phi^-_j \tilde{b}^*\tilde{t}}\,U^t \tau_3 U^{t\dagger}\Big)
 \nonumber\\
&&+\, \Big[\, B_0 (s,m^2_{\tilde{t}_1},m^2_{\tilde{b}_1})\: -\:
B_0 (s,m^2_{\tilde{t}_1},m^2_{\tilde{b}_2})\: -\:
B_0 (s,m^2_{\tilde{t}_2},m^2_{\tilde{b}_1})\: +\: 
B_0 (s,m^2_{\tilde{t}_2},m^2_{\tilde{b}_2})\,\Big]\nonumber\\
&&\times\, {\rm Tr}\, \Big( \Gamma^{\phi^+_i \tilde{t}^*\tilde{b}}\,
U^b \tau_3 U^{b\dagger}\, \Gamma^{\phi^-_j \tilde{b}^*\tilde{t}}\,
U^t \tau_3 U^{t\dagger} \Big)\, \bigg\}\, ,\\
  \label{Pichb}
\Pi^{\pm,(b)}_{ij} (s) & =& -\, \delta_{ij}\, 
N_c\, \frac{1}{32\pi^2}\, \sum_{\tilde{q} =
\tilde{t},\tilde{b}}\,
\bigg\{\, \Big[\, A_0 (m^2_{\tilde{q}_1})\: +\: A_0
(m^2_{\tilde{q}_2})\, \Big]\, {\rm Tr}\, 
\Big( \Gamma^{\phi^+_i\phi^-_i \tilde{q}^*\tilde{q}}\Big)\nonumber\\
&&+\, \Big[\, A_0 (m^2_{\tilde{q}_1})\: -\: A_0
(m^2_{\tilde{q}_2})\, \Big]\, 
{\rm Tr}\, \Big( 
\Gamma^{\phi^+_i\phi^-_i \tilde{q}^*\tilde{q}}\, 
U^q \tau_3 U^{q\dagger}\Big)\, \bigg\}\, .
\end{eqnarray}

In addition  to squark loops, top and  bottom quarks also contribute to 
the neutral  and  charged   Higgs-boson  self-energies.
The  $t$- and  $b$-quark   contributions to  the   CP-even
and CP-odd Higgs-boson self-energies are given explicitly by
\begin{eqnarray}
  \label{PiSf}
\Pi^{S,(c)}_{11} (s) & = & -\,N_c\, \frac{|h_b|^2}{8\pi^2}\, \Big[\,
A_0 (m^2_b)\: -\: \frac{1}{2}\, (s - 4m^2_b)\, B_0( s, m^2_b, m^2_b)\,
\Big],\nonumber\\
\Pi^{S,(c)}_{22} (s) & = & -\,N_c\, \frac{|h_t|^2}{8\pi^2}\, \Big[\,
A_0 (m^2_t)\: -\: \frac{1}{2}\, (s - 4m^2_t)\, B_0( s, m^2_t, m^2_t)\,
\Big], \nonumber\\
  \label{PiPf}
\Pi^{P,(c)}_{11} (s) & = & N_c\, \frac{|h_b|^2}{8\pi^2}\, \Big[\,
A_0 (m^2_b)\: -\: \frac{s}{2}\, B_0( s, m^2_b, m^2_b)\, \Big],\nonumber\\
\Pi^{P,(c)}_{22} (s) & = & N_c\, \frac{|h_t|^2}{8\pi^2}\, \Big[\,
A_0 (m^2_t)\: -\: \frac{s}{2}\, B_0( s, m^2_t, m^2_t)\, 
 \Big]\, ,
\end{eqnarray}
and  we note that $\Pi^{S,(c)}_{12} (s)  = \Pi^{P,(c)}_{12} (s) = 0$, and
also $\Pi^{SP,(c)}_{ij} (s) = 0$, for $i,j = 1,2$.

By analogy, the charged Higgs-boson self-energies also receive quantum
corrections due to $t$ and $b$ quarks, viz.
\begin{eqnarray}
  \label{Pplus}
\Pi^{\pm,(c)}_{11} (s) & =& -\, N_c\, \frac{|h_b|^2}{16\pi^2}\,
\Big[\, A_0(m^2_t)\: +\: A_0(m^2_b)\: - 
\: (s - m^2_t - m^2_b)\, B_0( s, m^2_t, m^2_b)\, \Big]\,,\nonumber\\
\Pi^{\pm,(c)}_{12} (s) & =& N_c\, \frac{|h_t h_b|}{8\pi^2}\,
m_t m_b\, B_0( s, m^2_b, m^2_t)\, ,\nonumber\\
\Pi^{\pm,(c)}_{22} (s) & =& -\, N_c\, \frac{|h_t|^2}{16\pi^2}\,
\Big[\, A_0(m^2_t)\: +\: A_0(m^2_b)\: -\: 
(s - m^2_t - m^2_b)\, B_0( s, m^2_t, m^2_b)\, \Big]\, .\qquad
\end{eqnarray}

The charged and neutral Higgs-boson self-energies derived here are not
UV-safe quantities.  As has been discussed in  Section 2,  they can be
renormalized in  the $\overline{\rm  MS}$  scheme by  neglecting terms
proportional to $C_{\rm UV}$. However, even in the $\overline{\rm MS}$
scheme,  one  still   has  to take   into  consideration the  relevant
Higgs-boson tadpole CTs.  Therefore,  we quote the analytic results of
the tadpole   parameters $T_{\phi_{1,2}}$  and   $T_{a_{1,2}}$.  After
taking  into account the   third-generation squark/quark  loop effects
displayed in Fig.~\ref{f1}(d) and (e), we obtain
\begin{eqnarray}
  \label{Tphi12}
T^{(d)}_{\phi_i} & = & N_c\, \frac{1}{32\pi^2}\, \sum_{\tilde{q} =
\tilde{t},\tilde{b}}\,
\bigg\{\, \Big[\, A_0 (m^2_{\tilde{q}_1})\: +\: A_0
(m^2_{\tilde{q}_2})\, \Big]\, {\rm Tr}\, 
\Big( \Gamma^{\phi_i \tilde{q}^*\tilde{q}}\Big)\nonumber\\
&&+\, \Big[\, A_0 (m^2_{\tilde{q}_1})\: -\: A_0
(m^2_{\tilde{q}_2})\, \Big]\, 
{\rm Tr}\, \Big( 
\Gamma^{\phi_i \tilde{q}^*\tilde{q}}\, U^q \tau_3 U^{q\dagger}
\Big)\, \bigg\}\, ,\nonumber\\
T^{(e)}_{\phi_1} & = & N_c\, \frac{v_1 |h_b|^2}{8\pi^2}\, 
                                        A_0 (m^2_b)\, ,\nonumber\\
T^{(e)}_{\phi_2} & = & N_c\, \frac{v_2 |h_t|^2}{8\pi^2}\, 
                                        A_0 (m^2_t)\, ,\nonumber\\
  \label{Ta12}
T^{(d)}_{a_1} & = & N_c\, \frac{1}{32\pi^2}\, \sum_{\tilde{q} =
\tilde{t},\tilde{b}}\, 
\Big[\, A_0 (m^2_{\tilde{q}_1})\: -\: A_0
(m^2_{\tilde{q}_2})\, \Big]\, {\rm Tr}\, \Big( 
\Gamma^{a_1 \tilde{q}^*\tilde{q}}\, 
U^q \tau_3 U^{q\dagger}\Big)\,,\nonumber\\
T^{(d)}_{a_2} & = & -\, \tan\beta\ T^{(d)}_{a_1}\, ,
\end{eqnarray}
with  $T_{\phi_{1,2}} =  T^{(d)}_{\phi_{1,2}} +  T^{(e)}_{\phi_{1,2}}$
and $T_{a_{1,2}} = T^{(d)}_{a_{1,2}}$.

\end{appendix}

%
%
\begin{figure}
   \leavevmode
 \begin{center}
   \epsfxsize=16.2cm
    \epsffile[0 0 539 652]{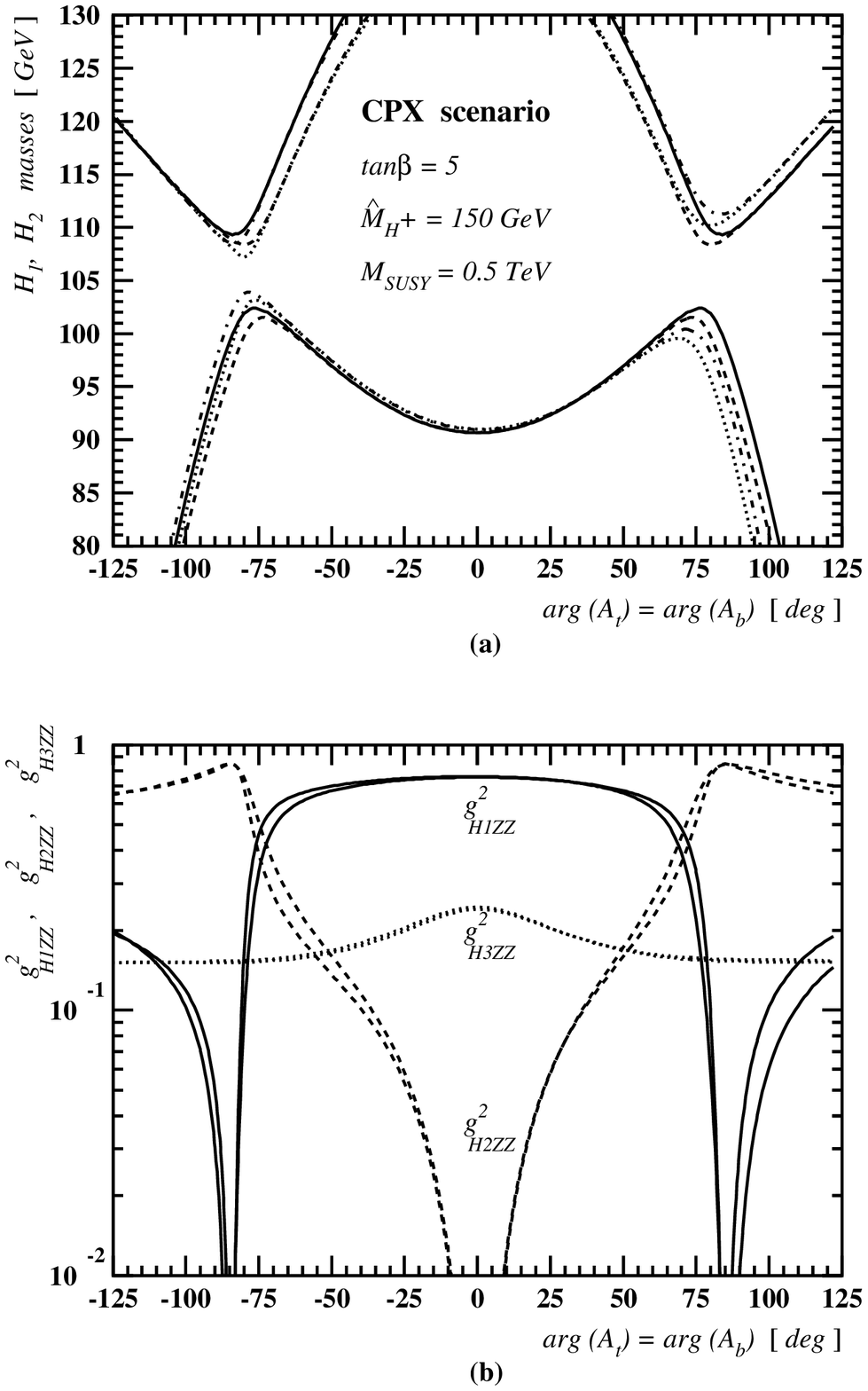}
 \end{center}
 \vspace{-2.cm} 
\caption{\it Numerical estimates of (a) the $H_{1,2}$-
effective-potential and pole masses and (b)~$g^2_{H_iZZ}$ as functions
of ${\rm arg}\, (A_t)$, in a CPX scenario with $M_{\rm SUSY} =
0.5$~TeV, ${\rm arg}\, (m_{\tilde{g}}) = 0$ and $90^\circ$. In
plot~(a), the effective-potential mass $M_{H_1}~(M_{H_2})$ is
indicated by a solid (dash-dotted) line for ${\rm arg}\,
(m_{\tilde{g}}) = 0~(90^\circ )$, and its pole mass
$\widehat{M}_{H_1}~(\widehat{M}_{H_2})$ by a dashed (dotted) line for
${\rm arg}\, (m_{\tilde{g}}) = 0~(90^\circ )$.}\label{fig:pole1}
\end{figure}

%
%
\begin{figure}
   \leavevmode
 \begin{center}
   \epsfxsize=16.2cm
    \epsffile[0 0 539 652]{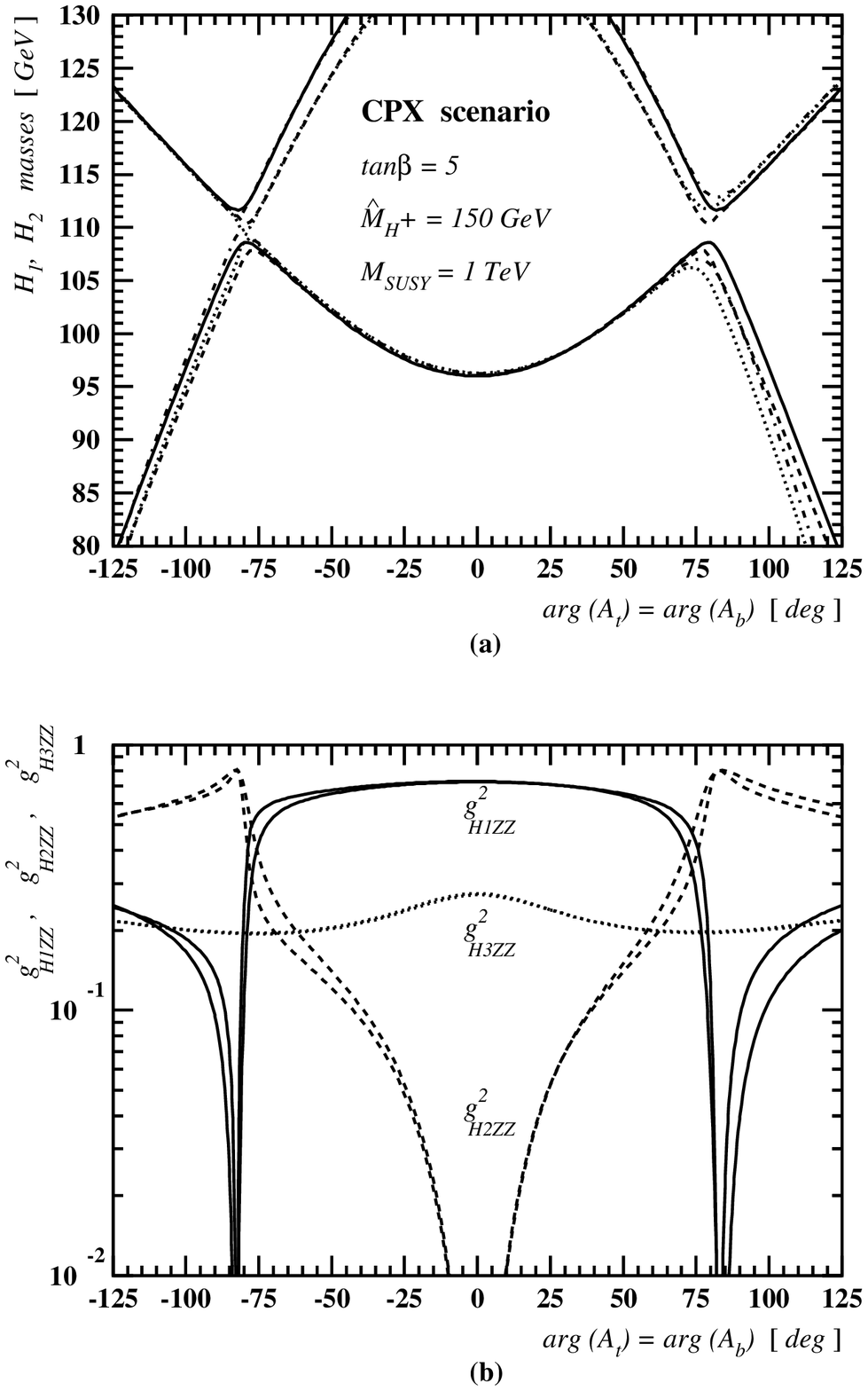}
 \end{center}
 \vspace{-1.cm} 
\caption{\it The same as in Fig.~\ref{fig:pole1}, but for a CPX
scenario with $M_{\rm SUSY} = 1$~TeV.}\label{fig:pole2}
\end{figure}

%
%
\begin{figure}
   \leavevmode
 \begin{center}
   \epsfxsize=16.2cm
    \epsffile[0 0 539 652]{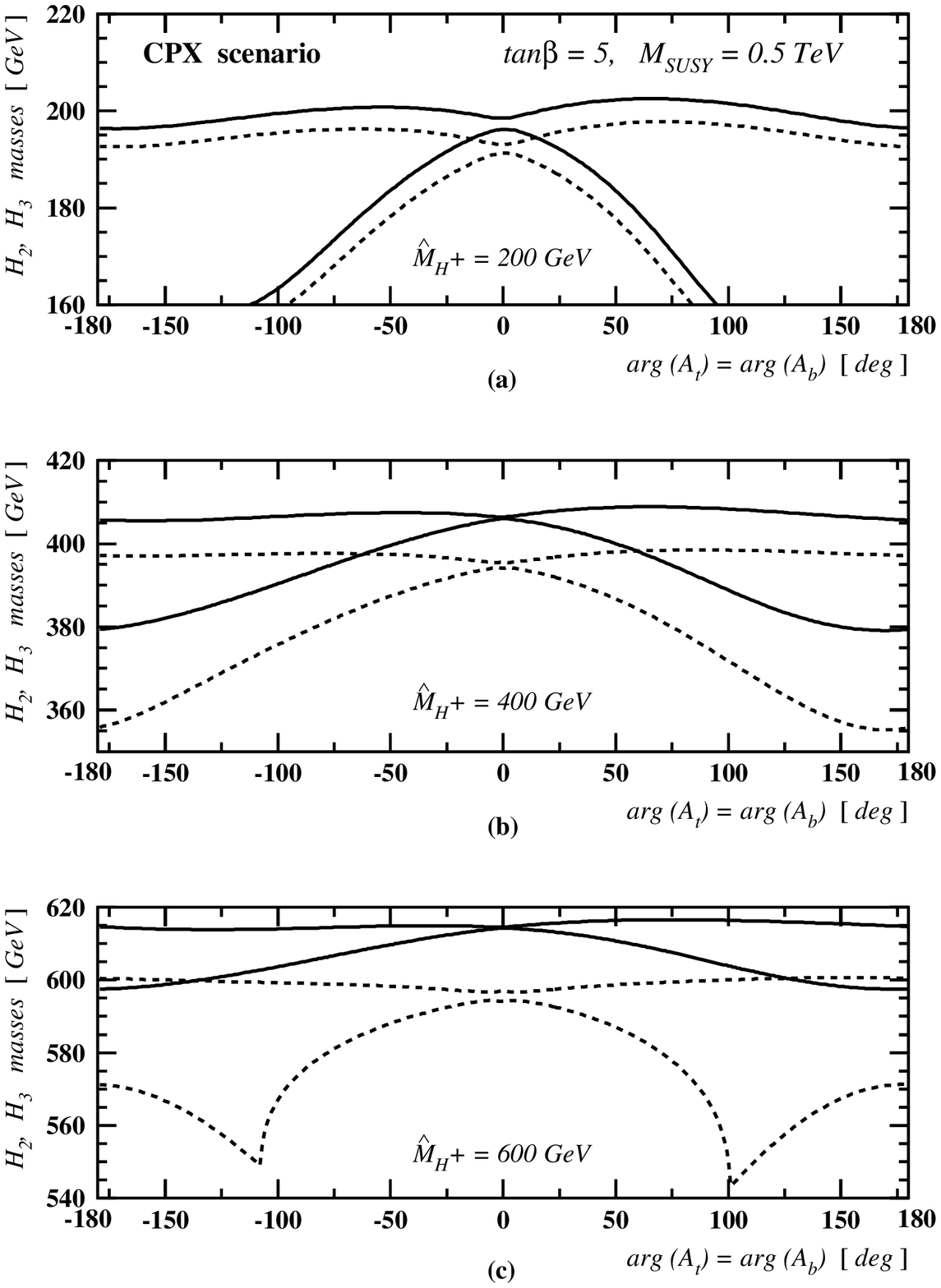}
 \end{center}
 \vspace{-1.cm} 
\caption{\it Numerical estimates of the two heaviest $H_2$- and
$H_3$-boson masses versus ${\rm arg}\, (A_t)$ for different charged
Higgs-boson pole masses in a CPX scenario with $arg ( m_{\tilde{g}} )
= 90^\circ$. Effective-potential masses are indicated by solid lines
and pole masses by dashed ones.}\label{fig:pole3}
\end{figure}

%
%
\begin{figure}
   \leavevmode
 \begin{center}
   \epsfxsize=16.2cm
    \epsffile[0 0 539 652]{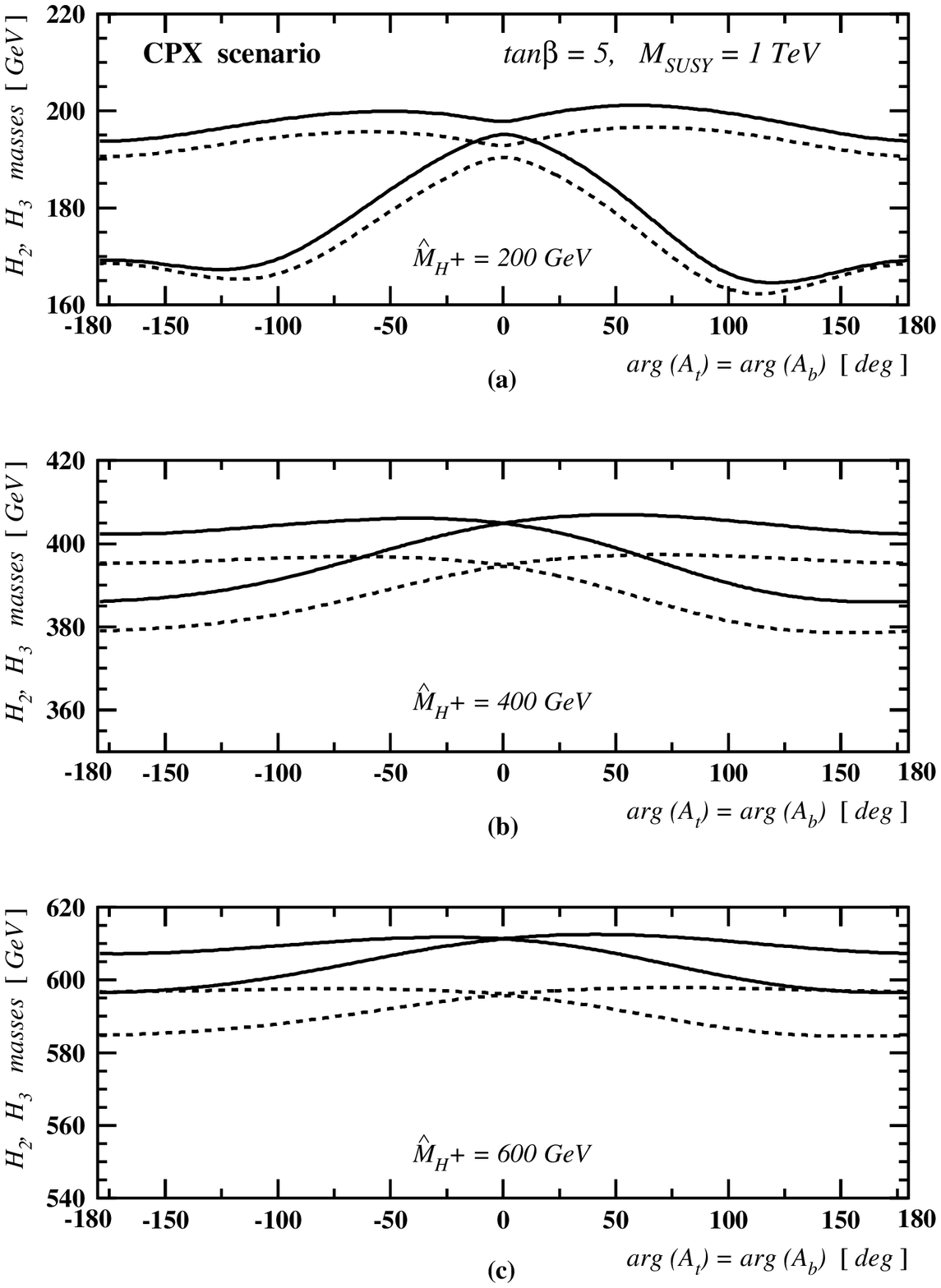}
 \end{center}
 \vspace{-1.cm} 
\caption{\it The same as in Fig.~\ref{fig:pole3}, but for a CPX
scenario with $M_{\rm SUSY} = 1$~TeV.}\label{fig:pole4}
\end{figure}

%
%
\begin{figure}
   \leavevmode
 \begin{center}
   \epsfxsize=16.2cm
    \epsffile[0 0 539 652]{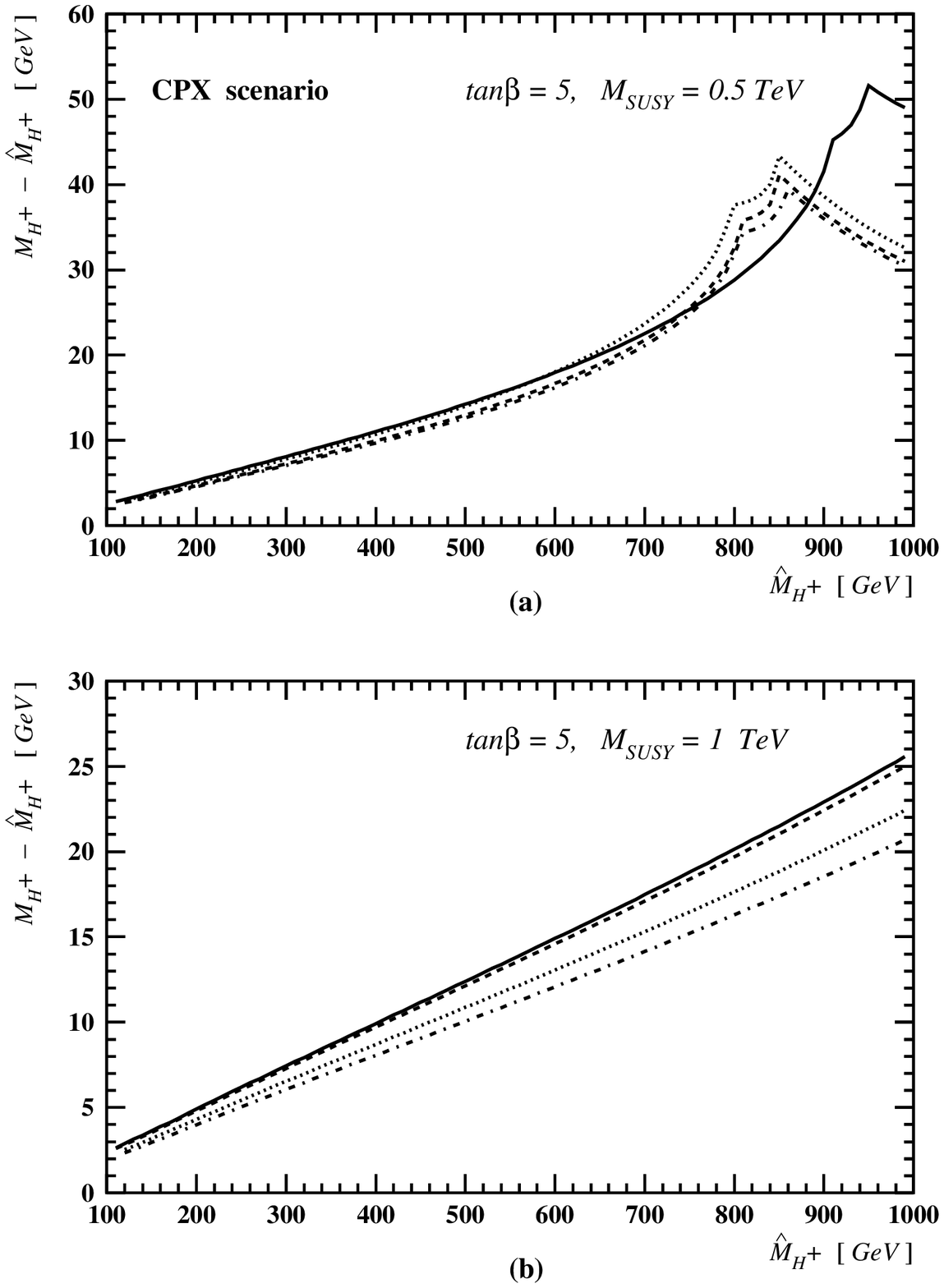}
 \end{center}
 \vspace{-2.cm} 
\caption{\it Numerical values of the difference between the
effective-potential and the pole mass of the charged Higgs boson,
$M_{H^+} - \widehat{M}_{H^+}$, as a function of $\widehat{M}_{H^+}$,
for a CPX scenario with (a) $M_{\rm SUSY} = 0.5$~TeV and (b) $M_{\rm
SUSY} = 1$~TeV. The different types of lines correspond to ${\rm
arg}\, (A_t) = {\rm arg}\, (m_{\tilde{g}}) = 0$ (solid), ${\rm arg}\,
(A_t) =90^\circ$ and ${\rm arg}\, (m_{\tilde{g}}) = 0$ (dashed), ${\rm
arg}\, (A_t) = {\rm arg}\, (m_{\tilde{g}}) = 90^\circ$ (dotted), and
${\rm arg}\, (A_t) =-90^\circ$ and ${\rm arg}\, (m_{\tilde{g}}) =
90^\circ$ (dash-dotted).}\label{fig:pole5}
\end{figure}

\end{document}